\documentclass[aps,prc,twocolumn,showpacs,amsmath,amssymb,nofootinbib]{revtex4-1}
\usepackage{graphicx}
\usepackage{color}
\usepackage{times}
\usepackage{inputenc}
\usepackage{bm}
\usepackage{ulem}
\usepackage{multirow}
\usepackage{float}
\usepackage{url}
\usepackage{natbib}
\usepackage{mathrsfs}
\usepackage{physics}
\usepackage[table,xcdraw]{xcolor}
\usepackage{booktabs}
\usepackage{comment}
\usepackage[colorlinks=true,citecolor=blue,urlcolor=blue,linkcolor=blue]{hyperref}
\begin{document}
\title{Pasta properties of the neutron star within effective relativistic mean-field model}
\author{Vishal Parmar$^{1}$}
\email{physics.vishal01@gmail.com}
\author{H. C. Das$^{2,3}$}
\email{harish.d@iopb.res.in}
\author{Ankit Kumar$^{2,3}$}
\email{ankit.k@iopb.res.in}
\author{Ankit Kumar$^{4}$}
\author{M. K. Sharma$^{1}$}
\author{P. Arumugam$^{4}$}
\author{S. K. Patra$^{2,3}$} 
\affiliation{\it $^{1}$ School of Physics and Materials Science, Thapar Institute of Engineering and Technology, Patiala 147004, India}
\affiliation{\it $^{2}$Institute of Physics, Sachivalaya Marg, Bhubaneswar 751005, India}
\affiliation{\it $^{3}$Homi Bhabha National Institute, Training School Complex, Anushakti Nagar, Mumbai 400094, India}
\affiliation{\it $^{4}$ Department of Physics, Indian Institute of Technology Roorkee, Roorkee 247667, Uttarakhand, India}
\date{\today}
\begin{abstract}
We study the properties of pasta structures and their influence on the neutron star observables employing the effective relativistic mean-field (E-RMF) model. The compressible liquid drop model is used to incorporate the finite size effects, considering the possibility of nonspherical structures in the inner crust. The unified equation of states are constructed for several E-RMF parameters to study various properties such as pasta mass and thickness in the neutron star's crust. The majority of the pasta properties are sensitive to the symmetry energy in the subsaturation density region. Using the results from Monte Carlo simulations, we estimate the shear modulus of the crust in the context of quasiperiodic oscillations from soft gamma-ray repeaters and calculate the frequency of fundamental torsional oscillation mode in the inner crust. Global properties of the neutron star such as mass-radius profile, the moment of inertia, crustal mass, crustal thickness, and fractional crustal moment of inertia are worked out. The results are consistent with various observational and theoretical constraints. 
\end{abstract}
\maketitle
\section{Introduction}
\label{intro}
In recent years, the origin, structure, and dynamics of neutron stars have played a central role in multimessenger and gravitational-wave astronomy \cite{Abbott_2017, Abbott_2018}. It has provided us with the opportunity to understand the behavior of fundamental forces in extreme environmental situations. With the available multifaceted data from various astrophysical observations,  we now can better constrain the neutron star observables such as mass, radius, tidal deformity, etc., and the behavior of the equation of state (EoS) over a wide density range \cite{Cromartie_2019, Antoniadis_2013, Demorest_2010}. The "crust" has drawn particular attention among various layers of its internal structure because of its complexity and importance in multiple astrophysical phenomena. This layer of the neutron star has a thickness $\approx$ 10 \% of the radius and mass $\approx$ 1 \% of the star's mass \cite{Haensel_2008}. Since a neutron star's crust contains nuclear matter at subsaturation density, it acts as the most advanced cosmic laboratory where theory can be confronted with neutron star observations. 

The crust is divided into two parts; the outer crust, which contains the ions arranged in a periodic lattice embedded in a strongly degenerate electron gas, and the inner crust, which has clusters of neutrons and protons. These clusters are surrounded by neutron gas along with the electron gas and are for the most part, an elastic solid \cite{Gearheart_2011} and body-centered cubic (BCC) type crystal \cite{Baiko_2011}. The composition of the outer crust can be estimated accurately up to some extent based on experimental atomic mass evaluations \cite{Huang_2021} along with the mass table from accurately calibrated models such as finite-range liquid-drop model (FRDM) \cite{MOLLER20161}, Hartree-Fock-Bogoliubov (HFB) \cite{hfb14, hfb2426} etc. using the Baym, Pethick and Sutherland technique (BPS) \cite{BPS_1971}. In contrast, the estimation of the composition of the inner crust is limited by the inevitability of using an empirical mass model because of our inability to measure the mass excess of highly neutron-rich nuclei in a neutron gas background. At low densities, the clusters are at a sufficient distance from each other and are expected to be spherical in shape \cite{Dinh_2021}. However, at high densities, i.e., near the crust-core transition density, the system becomes "frustrated" as a result of competition between the nuclear and Coulomb interactions \cite{Avancini_2008, Avancini_2009}. The frustration leads to the system arranging itself into various exotic geometries commonly known as "nuclear pasta" \cite{Ravenhall_1983,  Lorenz_1993, Maruyama_1998}. These configurations of nuclear pasta are related to the complex terrestrial fluids such as glassy system \cite{Newton_2022} rather than a solid and have a variety of responses towards the mechanical stimuli \cite{PETHICK19987, watanabe2007dynamical}. 

Although there exists no direct and robust observational evidence of nuclear pasta, various tantalizing observations indicate its existence \cite{Horowitz_2014,  Pons_2013, Horowitz_2015}. Numerous  theoretical attempts  based on molecular dynamics simulations \cite{Lin_2020, li2021tasting, watanabe2007dynamical}, compressible liquid-drop models \cite{Carreau_2019, Newton_2021}, Thomas-Fermi method \cite{Haensel_2008, Furtado_2021, bcpm} and nuclear density functional theory \cite{schuetrumpf2016clustering} point towards the possibility of the pasta structures near the crust-core transition density. The amount of these structures in the inner crust plays pivotal role in the explanation of  various neutron star mechanisms such as crust cooling \cite{Horowitz_2015, Ootes_2018}, spin period \cite{Pons_2013}, quasiperiodic oscillation in giant flares \cite{Steiner_2009}, transport \cite{rezzolla2018physics}, shattering of the crust \cite{Tsang_2012} etc.  The discovery of quasiperiodic oscillations (QPOs) in soft gamma repeaters (SGRs), which are related to the torsional vibrations of the neutron star crust, enables us to put constraints on the thickness and mass of the pasta structure and quadrupole ellipticity sustainable by the crust \cite{Gearheart_2011}. Theoretically, this is achieved by new approaches to nuclear models in the form of Bayesian inference \cite{Newton_2021, Carreau_2019, Balliet_2021}   and establishing possible correlations between parameters and crust properties through systematic surveys of models \cite{Newton_2013, Zhang_2018, Horowitz_2001, Oyamatsu_2007}. These approaches of nuclear models are essential to account for the constantly updating data on the nuclear matter and neutron star observables with improved quantity and fidelity. However, one must take a simplistic energy density functional to account for the computational requirements.    

In Ref. \cite{Parmar_2022}, we calculated three unified EoSs using the effective field theory motivated relativistic mean-field (E-RMF) model employing the widely used compressible liquid drop model (CLDM). We considered only spherical symmetric shapes in the inner crust to estimate various crustal properties of the neutron star. Since nonspherical configurations influence the microscopic properties of the neutron star, it is essential to have a unified treatment of EoS (same EoS from surface to the core) considering all the possible pasta structures. Therefore, to comprehensively understand the impact of pasta structure, we extend our calculations of \cite{Parmar_2022} for the case of nonspherical shapes. We consider 13 well-known parameter sets, namely,  
BKA24 \cite{Aggarwal_2010}, FSU2 \cite{Chen_2014}, FSUGarnet \cite{Chen_2015}, G1 \cite{Furnstahl_1997}, G2 \cite{Furnstahl_1997}, G3 \cite{Kumar_NPA_2017}, GL97 \cite{NKGb_1997}, IUFSU \cite{Fattoyev_2010}, IUFSU$^*$ \cite{Fattoyev_2010}, IOPB-I \cite{Kumar_2018}, SINPA \cite{Mondal_2016}, SINPB \cite{Mondal_2016} and TM1 \cite{Sugahara_1994}. Using these parameter sets, we construct the neutron star model by evaluating the unified EoS considering the existence of nonspherical shapes in the inner crust. In view of the recent Bayesian inference of crust properties, we calculate the mass and thickness of the pasta structures and investigate their dependency on the model used. The related properties such as the shear modulus of the crust and the frequency of fundamental torsional oscillation mode in context to the SGRs are also investigated. Finally, we calculate the global properties of the neutron star from the unified EoSs such as mass-radius ($M-R$) profile, total crust mass ($M_{\rm crust}$), and thickness ($l_{\rm crust}$), moment of inertia ($I$), fractional moment of inertia ($I_{\rm crust}/I$), etc.

The paper is organized as follows: In Sec. \ref{formulation}, we briefly describe the numerical procedure for calculating the composition of the outer and inner crust. We discuss the main ingredient of the CLDM and E-RMF formalism in \ref{cldmform} and \ref{ermfform},  emphasizing the inclusion of nonspherical structure or "nuclear pasta." The amount and thickness of various pasta structures are discussed in \ref{relativepasta}, shear modulus and velocity in \ref{shearmodulus} and an accurate description of neutron star observables in \ref{form:NS_observable}. The results are discussed in Sec. \ref{results}, and finally, we summarize our results in Section \ref{conclusion}. 

\section{Formalism}
\label{formulation}
We begin our calculations by using the pioneering variational formalism originally proposed by Baym, Pethick, and Sutherland (BPS) \cite{BPS_1971} to find the composition of the outer crust of the neutron star. We minimize the Gibbs free energy at fixed pressure \cite{Parmar_2022, Carreau_2019} where the atomic mass table serves as an input. We use the most recent AME2020 data \cite{Huang_2021} along with the recently measured mass excess of $^{77-79}$Cu taken from \cite{welker2017}, $^{82}$Zn from \cite{wolf2013} and $^{151-157}$Yb \cite{Beck_2021}  for the known masses and extrapolated them using the microscopic Hartree-Fock-Bogoliubov (HFB-26)  data which is based on the accurately calibrated Brussels-Montreal functional \cite{hfb2426}. To model the inner crust, we employ the famous CLDM used extensively in recent times for various problems of neutron star crust. We here discuss the model's main ingredient, emphasizing the inclusion of nonspherical structure or "nuclear pasta".


\subsection{ CLDM for nuclear pasta}
\label{cldmform}
The CLDM formulation originally proposed by Baym, Bethe, Pethick (BBP) \cite{Baym_1971} assumes a repeating unit cell of volume $V_{WS}$ in which clustered structure "pasta" resides, immersed in a uniform neutron gas of density $\rho_g$. The system is neutralized by a homogeneous ultra-relativistic electron gas of density $\rho_e$. Using the Wigner-Seitz (WS) approximation, the energy of the system in the inner crust of a neutron star can then be written as \cite{Newton_2013},
\begin{align}
    E(r_c, y_p,\rho, \rho_n)&=f(u)\left[E_{\rm bulk}(\rho_b, y_p)\right]
    \nonumber\\
    &
    +E_{\rm bulk}(\rho_g,0)\left[1-f(u)\right]
    \nonumber\\
    &
    +E_{\rm surf}+E_{\rm curv}+E_{\rm coul}+E_e.
    \label{eq:energy}
\end{align}
Here $r_c$ is the radius (half-width in the case
of planar geometry) of WS cell, $y_p$ the proton fraction, and $\rho$ and $\rho_n$ are the baryon density of charged nuclear component and density of neutron gas, respectively. The cluster is characterised by density $\rho_i$ and volume fraction $u$ as \cite{Newton_2012, Dinh_2021}
\begin{equation}
    u=\begin{cases}
       (\rho-\rho_g)/(\rho_i-\rho_g) \, \, \, \text{for clusters},\\ (\rho_i-\rho)/(\rho_i-\rho_g) \, \, \, \, \text{for holes}.
      \end{cases}
\end{equation}
The function $f(u)$ is defined as 
\begin{equation}
    f(u) = \begin{cases}
           u  \, \, \, \, \,  \text{for clusters}, \\
           1-u \, \, \, \text{for holes.}
          \end{cases}
\end{equation}
Pasta structure only affects the finite size effects, which can be expressed analytically as a function of the dimension of the pasta structure. We consider the three canonical geometries, namely spherical, cylindrical, and planar, defined by a dimensionality parameter $d = 3, 2, 1,$ respectively. We then define the finite size corrections along the same lines as in \cite{Newton_2013, Dinh_2021}. The surface and curvature energies are written as \cite{Newton_2013, Dinh_2021},
\begin{equation}
    E_{\rm surf}+E_{\rm curv}=\frac{u d}{r_N}\left( \sigma_s +\frac{(d-1)\sigma_c}{r_N}\right),
\end{equation}
where $r_N$ is the radius/half-width of the cluster/hole and   $\sigma_s$ and $\sigma_c$ are the dimension independent surface and curvature tension  based on the Thomas-Fermi calculations and are defined as \cite{Ravenhall_1983} 
\begin{equation}
\label{surf}
    \sigma_s=\sigma_0\frac{2^{p+1} + b_s}{y_p^{-p} +b_s+(1-y_p)^{-p}},
\end{equation}
\begin{equation}
\label{curv}
    \sigma_c=\alpha \, \sigma_s\frac{\sigma_{0,c}}{\sigma_0}\left(\beta-y_p\right).
\end{equation}
Here the parameters ($\sigma_0$, $\sigma_c$, $b_s$, $\alpha$, $\beta$, $p$) are optimised for a given equation of state on the atomic mass evaluation 2020 data \cite{Huang_2021}. The Coulomb energy reads as \cite{Dinh_2021} 
\begin{equation}
    E_{\rm coul}=2\pi (e\,y_p\,n_i\,r_N)^2\,u\,\eta_d(u),
\end{equation}
where e is the elementary charge and $\eta_d(u)$ is associated with the pasta structure as \cite{Dinh_2021, Newton_2013}
\begin{equation}
    \eta_d(u)=\frac{1}{d+2}\Big[\frac{2}{d-2}\Big(1-\frac{du^{1-\frac{2}{d}}}{2}\Big)+u\Big]
\end{equation}
for $d=1$ and 3 whereas for $d=2$ it reads as,
\begin{equation}
    \eta_d(u)=\frac{1}{4}\Big[\log (\frac{1}{u}) + u -1\Big].
\end{equation}
For a given baryon density, the equilibrium composition of a WS cell is obtained by minimizing the energy per unit volume using the variational method where the auxiliary function to be minimized reads as \cite{Parmar_2022, Carreau_2019}
\begin{equation}
    \mathcal{F}=\frac{E_{WS}}{V_{WS}}-\mu_b\rho.
\end{equation}
Here, $\mu_b$ is the baryonic chemical potential. This results in a set of four differential equations corresponding to mechanical, dynamical, $\beta$-equilibrium, and the nuclear virial theorem \cite{Carreau_2020, Carreau_2019}. The viral relation is used to numerically solve the value of $r_N$. To obtain the most stable pasta structure at a given baryon density, we first calculate the composition of a spherical nucleus. Then keeping this composition fixed, we calculate the radius or half-width of five different pasta structures, namely, sphere, rod, plate, tube, and bubble. The equilibrium phase is then the one that minimizes the total energy of the system. 
\subsection{Effective relativistic mean-field theory}
\label{ermfform}
The E-RMF formalism is based on an effective field theory motivated relativistic mean field model. This framework is consistent with the underlying Quantum chromodynamics symmetries and takes care of the renormalization problem in RMF theory. This formalism has been applied in a wide range of nuclear physics problems in the past few years \cite{MULLER_1996, Wang_2000, DelPairing_2001, Kumar_2020, Das_2020, Das_2021}. The E-RMF effective Lagrangian which include the interaction between different mesons, such as, $\sigma$, $\omega$, $\rho $, $\delta$ and photon is written as \cite{Patra_2002, Kumar_2017, Kumar_2018, Vishal_2021_jpg, Vishal_2021, DasBig_2020}, 
\begin{widetext}
\begin{align}
\label{rmftlagrangian}
\mathcal{E}(r)=&\psi^{\dagger}(r)\qty{i\alpha\cdot\grad+\beta[M-\Phi(r)-\tau_3D(r)]+W(r)+\frac{1}{2}\tau_3R(r)+\frac{1+\tau_3}{2} A(r)-\frac{i\beta \alpha }{2M}\qty(f_\omega \grad W(r)+\frac{1}{2}f_\rho \tau_3 \grad R(r))}\psi(r) \nonumber \\
&
+\qty(\frac{1}{2}+\frac{k_3\Phi(r)}{3!M}+\frac{k_4}{4!}\frac{\Phi^2(r)}{M^2})\frac{m^2_s}{g^2_s}\Phi(r)^2+\frac{1}{2g^2_s}\qty\Big(1+\alpha_1\frac{\Phi(r)}{M})(\grad \Phi(r))^2-\frac{1}{2g^2_\omega}\qty\Big(1+\alpha_2\frac{\Phi(r)}{M})(\grad W(r))^2 \nonumber\\
&
-\frac{1}{2}\qty\Big(1+\eta_1\frac{\Phi(r)}{M}+\frac{\eta_2}{2}\frac{\Phi^2(r)}{M^2})\frac{m^2_\omega}{g^2_\omega}W^2(r)-\frac{1}{2e^2}(\grad A^2(r))^2 -\frac{1}{2g^2_\rho}(\grad R(r))^2
-\frac{1}{2}\qty\Big(1+\eta_\rho\frac{\Phi(r)}{M})\frac{m^2_\rho}{g^2_\rho}R^2(r)
\nonumber \\
&
-\frac{\zeta_0}{4!}\frac{1}{g^2_\omega}W(r)^4-\Lambda_\omega(R^2(r)W^2(r))
+\frac{1}{2g^2_\delta}(\grad D(r))^2
+\frac{1}{2}\frac{m^2_\delta}{g^2_\delta}(D(r))^2.
\end{align} 
\end{widetext}
Here $\Phi(r)$, $W(r)$, $R(r)$, $D(r)$ and $A(r)$ are the fields corresponding to $\sigma$, $\omega$, $\rho$ and  $\delta $ mesons and photon respectively. The $g_s$, $g_{\omega}$, $g_{\rho}$, $g_{\delta}$ and $\frac{e^2}{4\pi }$  are the corresponding coupling constants and $m_s$, $m_{\omega}$, $m_{\rho}$ and $m_{\delta}$ are the corresponding masses. The zeroth component $T_{00}= H$ and the third component $T_{ii}$ of energy-momentum tensor 
\begin{equation}
\label{set}
T_{\mu\nu}=\partial^\nu\phi(x))\frac{\partial\mathcal{E}}{\partial\partial_\mu \phi(x)}-\eta^{\nu\mu}\mathcal{E},
\end{equation}
yields the energy and pressure density. The details regarding the equation of motion, chemical potential,  and effective mass can be found in \cite{Kumar_2018, Das_2019, Dutra_2016}. 

\subsection{Relative pasta layer thickness and mass}
\label{relativepasta}
It is shown in Refs. \cite{Lattimer_2007, Zdunik_2017} that the relative thickness and mass of the crust are correlated with mass, radius, and a single parameter of the core-crust interface, which depends on the EoS. In the same line, Newton {\it et al.} \cite{Newton_2021} derived the relative thickness and mass of a single layer of pasta structure as,
\begin{equation}
\label{eq:rr}
    \frac{\Delta R_p}{\Delta R_c} \approx \frac{\mu_c-\mu_p}{\mu_c-\mu_0},
\end{equation}
\begin{equation}
\label{eq:pp}
    \frac{\Delta M_p}{\Delta M_c} \approx 1-\frac{P_p}{P_c}.
\end{equation}
Here, $\mu_c$, $\mu_p$, and $\mu_0$ are the baryon chemical potential at crust-core (CC) transition, the location at which the pasta structure starts and at the surface of the star. $P_p$ and $P_c$ are the pressure at the bottom of the pasta layer and at the CC transition. Further, since the moment of inertia ($I$) of the crust is directly proportional to the mass of the crust to the first order of approximation \cite{Lorenz_1993}, therefore
\begin{equation}
    \frac{\Delta M_p}{\Delta M_c} \approx  \frac{\Delta I_p}{\Delta I_c}. 
\end{equation}
\subsection{Shear modulus and velocity}
\label{shearmodulus}
The shear modulus ($\mu$) of a BCC Coulomb lattice in a uniform electronic background (using the low-temperature limit) and  including electron screening effects as per within the Monte Carlo simulation \cite{Chugnov_2010} can be written as \cite{tews_2017, Sotani_2013},
\begin{equation}
\label{eq:shearmodulus}
    \mu=0.1194\left(1-0.010Z^{2/3}\right)\frac{\rho_i\left(Ze\right)^2}{a}.
\end{equation}
Here, $\rho_i$ is the density of nuclei, $Ze$ the charge and $a=R_{WS}$. Eq. (\ref{eq:shearmodulus}) is applicable for the case of spherical nuclei, whereas, near the crust-core boundary, there is a possibility of stable pasta structures. Although the exact elastic nature of these "exotic structures" is still unknown, one expects a decrease in the rigidity of the crust. To model this behavior, i.e., between the density region $\rho_{ph} \le \rho_b \le \rho_c$, where $\rho_{ph}$ and $\rho_c$ are the density at which nonspherical shapes appear and crust core transition density, respectively, we use a function which joins these regions smoothly as \cite{Sotani_2012, Gearheart_2011, Passamonti_2016}
\begin{equation}
\label{eq:mubar}
    \Bar{\mu}=c_1\left(\rho_b-\rho_c\right)\left(\rho-c_2\right),
\end{equation}
where $c_1$ and $c_2$ are the constants determined from  the boundary condition that $\Bar{\mu}$ should connect with Eq. (\ref{eq:shearmodulus}) smoothly at $\rho_b=\rho_{ph}$ and become zero smoothly at crust-core boundary. The latter condition arises from the fact that shear speed becomes zero at the crust-core boundary. We then define the shear speed as \cite{tews_2017},
\begin{equation}
\label{eq:shearspeed}
    V_s=\sqrt{\frac{\mu}{\rho_d}},
\end{equation}
with $\rho_d$ being the dynamical mass density. Neglecting the effects of neutron superfluidity, the dynamical mass density equals the total mass density (i.e. $\rho_d=\rho_m$)  \cite{Steiner_2009}. The frequency of the fundamental torsional oscillation mode can  be estimated from the plane wave analysis of the crustal shear perturbation equation \cite{Piro_2005} and is written as \cite{Samuelsson_2006, Gearheart_2011}
\begin{equation}
\label{eq:freq}
    \omega_0^2 \approx \frac{e^{2\nu} V_s^2 (l-1)(l+2)}{2RR_c},
\end{equation}
where $e^{2\nu}= 1-2M/R$, $R$ is the radius of the star, $R_c$ is the radius of the crust and  $l$  is the angular 'quantum' number.
\subsection{Neutron star observables}
\label{form:NS_observable}
For a static star, the macroscopic properties such as $M$ and $R$ of the neutron star can be found  by solving the Tolmann-Oppenheimer-Volkoff (TOV) equations as follow \cite{TOV1, TOV2}
\begin{eqnarray}
\frac{dP(r)}{dr}= - \frac{[P(r)+{\cal{E}}(r)][m(r)+4\pi r^3 P(r)]}{r[r-2m(r)]},
\label{eq:pr}
\end{eqnarray}
and 
\begin{eqnarray}
\frac{dm(r)}{dr}=4\pi r^2 {\cal{E}}(r).
\label{eq:mr}
\end{eqnarray}
The $M$ and $R$ of the star can be calculated with boundary conditions $r=0, P = P_c$ and $r=R, P = P_0$ at certain central density.

The moment of inertia (MI) of the neutron star is calculated in the Refs. \cite{Stergioulas_2003,Jha_2008,Sharma_2009,Friedmanstergioulas_2013,Paschalidis_2017,Quddus_2019,Koliogiannis_2020}. The expression of $I$ of uniformly rotating neutron star with angular frequency $\omega$ is given as \cite{Hartle_1967,Lattimer_2000,Worley_2008}
\begin{equation}
I \approx \frac{8\pi}{3}\int_{0}^{R}\ dr \ ({\cal E}+P)\  e^{-\phi(r)}\Big[1-\frac{2m(r)}{r}\Big]^{-1}\frac{\Bar{\omega}}{\Omega}\ r^4,
\label{eq:moi}
\end{equation}
where $\Bar{\omega}$ is the dragging angular velocity for a uniformly rotating star. The $\Bar{\omega}$ satisfying the boundary conditions are 
\begin{equation}
\Bar{\omega}(r=R)=1-\frac{2I}{R^3},\qquad \frac{d\Bar{\omega}}{dr}\Big|_{r=0}=0 .
\label{eq:omegabar}
\end{equation}
In order to calculate the accurate core/crust thickness or mass, one needs to integrate the TOV Eqs. \ref{eq:pr} and \ref{eq:mr} from $R=0$ to $R=R_{\rm core}$, which depends on pressure as $P(R=R_{\rm core})=P_t$. We calculate the crustal MI by using the Eq. (\ref{eq:moi}) from transition radius ($R_c$) to the surface of the star ($R$), which is given by \cite{Fattoyev_2010, Basu_2018} 
\begin{equation}
I_{\rm crust} \approx \frac{8\pi}{3}\int_{R_c}^{R}\ dr\ ({\cal E}+P)\  e^{-\phi(r)}\Big[1-\frac{2m(r)}{r}\Big]^{-1}\frac{\Bar{\omega}}{\Omega}\ r^4.
\label{eq:moic}
\end{equation}
\section{Results and Discussions}
\label{results} 
In this work, we use the CLDM to calculate the finite-size effects such as surface, curvature, Coulomb, etc. This method has been widely used to calculate the structure of the crust and other crustal properties such as pairing, thermal, entrainment properties, etc. \cite{BONCHE1981496, CHAMEL2006263, CHAMEL2005109}. In literature, the structure of neutron star crust has also been calculated using the well-known self-consistent Thomas-Fermi model, where energy is a function of density. The Thomas-Fermi calculations are carried out using either the WS approximation \cite{Avancini_2008, Maruyama_2005} where only typical pasta structures such as the sphere, rod, tube slab, and bubble are considered or using periodic boundary condition assuming no geometrical symmetry \cite{Minoru_2013}. In these calculations, the surface and Coulomb effects are calculated in a self-consistent manner and therefore are expected to give a better description of neutron star crust observables, considering the sensitivity to the very small energy difference between various pasta structures \cite{Pearson_2018}. However,  the solution of the self-consistent coupled equations in the Thomas-Fermi method is quite complicated and suffers from various technical difficulties such as boundary value problems \cite{Avancini_2008}. Such calculations, especially those considering no general geometrical shapes, are computationally expensive. On the other hand, the CLDM approach has the advantage that a proper description of surface and curvature effects estimates the crust properties at par with the Thomas-Fermi calculation \cite{Newton_2013} and  Hartree-Fock+BCS calculations \cite{Newton_2022}. The CLDM approach is computationally fast and thus works as a useful tool in the much-needed Bayesian Inference study of neutron star crust observables \cite{Newton_2021, Balliet_2021}. This method also gives a good treatment of the warm nuclear matter below sub-saturation density \cite{carreau2020crystallization}. The CLDM method, however, is limited by the fact that the WS cell is considered a body center cubic (BCC) type. At the same time, some Thomas-Fermi calculations predict the appearance of
face-centered cubic (fcc) lattice of droplets \cite{Xia_2021, Minoru_2013} in the inner crust of the neutron star.

For the equation of state (EoS), we use thirteen effective relativistic mean-field parameter sets to investigate the influence of pasta structures on neutron star properties. We show the saturation properties of the parameter sets in Table \ref{tab:forceproperties} along with the available empirical/experimental values. The motivation of taking these parameter sets lies in the fact that these sets  are the only few among hundreds of relativistic parameters \cite{Dutra_2016}, that reasonably satisfy  the observational constraints from  different massive pulsars such as PSR J0348+0432 ($M = 2.01\pm{0.04} \ M_\odot$) \cite{Antoniadis_2013}, PSR J0740+6620 ($M = 2.14_{-0.09}^{+0.10} \ M_\odot$) \cite{Cromartie_2019} and the radius constraints given by Miller {\it et al.} \cite{Miller_2019}, Riley {\it et al.} \cite{Riley_2019} and PSR J0030+0451 with X-ray Multi-Mirror Newton for canonical star with $R_{1.4} = 12.35 \pm 0.75$ km \cite{Miller_2021}. In addition, these sets also reproduce the finite nuclear properties at par with the experimental values and abide by the relevant nuclear matter constraints on EoS such as flow and kaon data, isoscalar giant monopole resonance, etc. \cite{Dutra_2014}. These sets are differentiated from each other by a wide range of saturation properties and various mesons self and cross-couplings.
\begin{table*}
\centering
\caption{Saturation properties of nuclear matter such as saturation density ($\rho_{\rm sat}$), binding energy ($B/A$), effective mass ($M^*/M$), incompressibility ($K$), symmetry energy ($J$, $J^{0.05}$ ), slope parameter ($L$, $L^{0.05}$) at saturation density and at $\rho=0.05$ fm$^{-3}$, curvature  of symmetry energy ($K_{\rm sym}$)  of nuclear matter for 13 relativistic parameter sets.}
\label{tab:forceproperties}
\renewcommand{\arraystretch}{1.0}
\begin{tabular}{lllllllllll}
\hline \hline
\begin{tabular}[c]{@{}l@{}} Parameter \\ sets \end{tabular}  &\hspace{0.1cm} $\rho_{\rm sat}$ &\hspace{0.1cm} $B/A$ & $M^*/M$ & \hspace{0.1cm}$K$ &\hspace{0.1cm} $J$ & \hspace{0.1cm}$L$ &\hspace{0.1cm} $K_{sym}$& \hspace{0.1cm} $J^{0.05}$ & \hspace{0.1cm} $L^{0.05}$&\hspace{0.1cm} $\Delta r_{np}^{^{208}Pb}$ \\ \hline 
BKA24 \cite{Aggarwal_2010} & 0.147 & -15.95 & 0.600 & 227.06 & 34.19 & 84.80 & -14.95& 14.53 & 33.88 & 0.240\\ \hline
FSU2 \cite{Chen_2014}& 0.150 & -16.28 & 0.593 & 238.00 & 37.62 & 112.80 & -24.25& 13.16 & 35.72 &0.287 \\ \hline
FSUGarnet \cite{Chen_2015} & 0.153 & -16.23 & 0.578 & 229.50 & 30.95 & 51.04 & 59.36& 18.07 & 32.11 &0.162 \\ \hline
G1 \cite{Furnstahl_1997}& 0.153 & -16.14 & 0.634 & 215.00 & 38.50 & 123.19 & 96.87& 12.96 & 35.51 &0.281\\ \hline
G2 \cite{Furnstahl_1997}& 0.154 & -16.07 & 0.664 & 215.00 & 36.40 & 100.67 & -7.28& 13.3 & 34.81 &0.256 \\ \hline
G3 \cite{Kumar_NPA_2017}& 0.148 & -16.02 & 0.699 & 243.96 & 31.84 & 49.31 & -106.07& 15.66 & 36.78 &0.180 \\ \hline
GL97 \cite{NKGb_1997}& 0.153 & -16.30 & 0.780 & 240.00 & 32.50 & 89.40 & -6.37& 11.95 & 31.00 & ------\\ \hline
IUFSU \cite{Fattoyev_2010}& 0.155 & -16.40 & 0.670 & 231.33 & 31.30 & 47.21 & 28.53& 17.80 & 33.85 &0.160 \\ \hline
IUFSU* \cite{Fattoyev_2010} & 0.150 & -16.10 & 0.589 & 236.00 & 29.85 & 51.508 & 7.87& 15.73 & 32.26 &0.164 \\ \hline
IOPB-I \cite{Kumar_2018}& 0.149 & -16.10 & 0.650 & 222.65 & 33.30 & 63.58 & -37.09 & 15.60 & 37.2 &0.221 \\ \hline
SINPA \cite{Mondal_2016}& 0.151 & -16.00 & 0.580 & 203.00 & 31.20 & 53.86 & -26.75& 17.02 & 33.59 &0.183 \\ \hline
SINPB \cite{Mondal_2016}& 0.150 & -16.04 & 0.634 & 206.00 & 33.95 & 71.55 & -50.57& 14.98 & 36.70 &0.241 \\ \hline
TM1 \cite{Sugahara_1994}& 0.145 & -16.30 & 0.634 & 281.00 & 36.94 & 111.00 & 34.00& 13.45 & 36.47 & 0.271\\ 
\hline
EMP/EXP  & 0.148/0.185    & -15.0/-17.0   & 0.55/0.6 &  220/260  &30.0/33.70 & 35.0/70.0   & -174.0/31.0 &-------&-------&0.212/0.354 \\
 &   \cite{bethe}  &    \cite{bethe} &  \cite{marketin2007}&   \cite{GARG201855} & \cite{DANIELEWICZ20141}&  \cite{DANIELEWICZ20141} &  \cite{zimmerman2020measuring} &&& \cite{Adhikari_2021}  \\
\hline
\hline
\end{tabular}%
\end{table*}

Among the parameter sets, GL97 \cite{NKGb_1997} contains only the nonlinear self couplings ($k_3$ and $k_4$) of $\sigma$ mesons, which reduces the incompressibility at par with the excepted range \cite{DANIELEWICZ20141}. TM1 \cite{Sugahara_1994} set takes into account the self-coupling of $\omega$-meson ($\zeta_0$) to soften the EoS at higher density. Parameter sets FSU2 \cite{Chen_2014}, IUFSU \cite{Fattoyev_2010}, IUFSU$^*$ \cite{Fattoyev_2010}, SINPA \cite{Mondal_2016}, SINPB \cite{Mondal_2016} incorporate the cross-coupling ($\Lambda_\omega$) between $\rho-\omega$ meson which helps in better agreement with the skin thickness ($r_n-r_p$)  and the symmetry energy data \cite{Todd_2003, Todd_2005}. The parameter sets based on E-RMF such as G1 and G2 \cite{Furnstahl_1997} consider the cross-couplings  $\eta_1$, $\eta_2$ and $\eta_\rho$ while excluding $\Lambda_\omega$. These sets give a soft EoS consistent with the koan and flow data \cite{Arumugam_2004}. In the line of E-RMF, recent forces FSUGarnet \cite{Chen_2015}, IOPB-I \cite{Kumar_2018} and G3 \cite{Kumar_NPA_2017} are designed for the calculation of finite nuclei and neutron star properties. G3 set contains all the couplings present in Eq. (\ref{rmftlagrangian}) and has an additional $\delta$ meson which is an important ingredient in the high-density regime \cite{Singh_2014}. All these forces are extensively used in the literature to estimate various nuclear matter properties ranging from nuclear reaction to nuclear structure and neutron star properties. In this work, we use these relativistic models to comprehensively study the crust properties of a neutron star and the influence of pasta structures on it. 

We begin our calculations from the surface of the neutron star using the formalism given in Section \ref{formulation} and calculate the outer crust EoS. Then for every model, as shown in Table \ref{tab:forceproperties}, we calculate the inner crust EoS using the CLDM  formalism considering all the available pasta structures. We discuss them in the following section. 
\subsection{Pasta phase within CLDM approximation}
\begin{figure}
    \centering
    \includegraphics[width=0.45\textwidth]{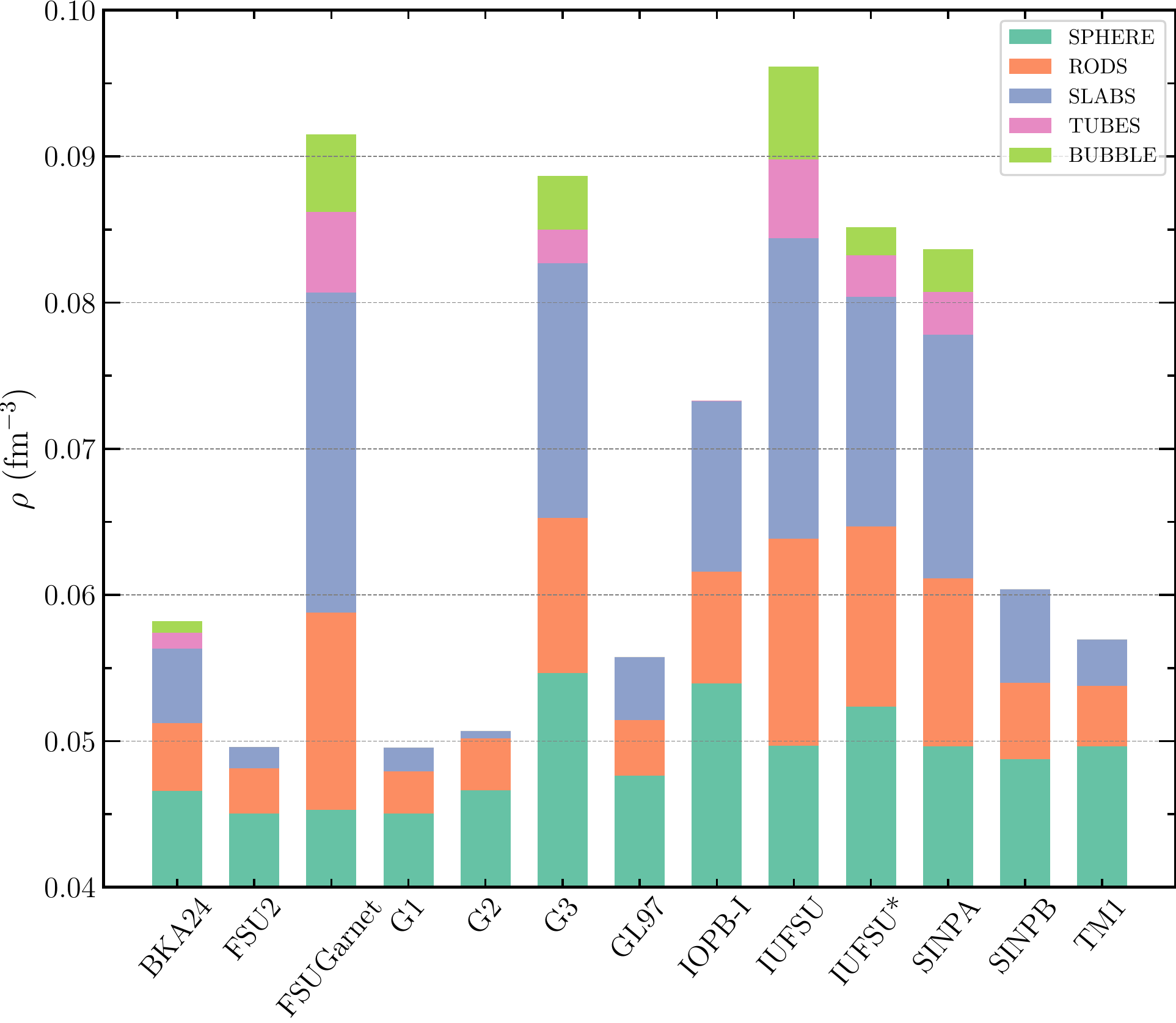}
    \caption{Comparison of the sequence of ground state pasta phase appearance for various functional.}
    \label{fig:past_bar}
\end{figure}

We present the result of our calculations for the pasta phase in the inner crust of the neutron star using various relativistic parameters in Fig. \ref{fig:past_bar} using the CLDM approximation. \textcolor{blue}{} Different colors represent the density regions where different pasta structures dominate. The edge in each bar represents the transition density of inhomogeneous crust to liquid homogeneous core. It is seen that the spherical geometry dominates for the majority of the inner crust extending up to $\rho \approx 0.05$ fm$^{-3}$ from the outer crust boundary, which is in agreement with various semi-classical and microscopic calculations \cite{Pearson_2018, Pearson_2020, BKS_2015}. There are two categories of parameter sets; one (FSU2, G1, G2, GL97, IOPB-I, SINPB, TM1) that estimates the pasta structure sequence as spheres $\rightarrow$ rods $\rightarrow$ slabs, and second (BKA24, FSUGarnet, G3, IUFSU, IUFSU$^*$, SINPA) that follow spheres $\rightarrow$ rods $\rightarrow$ slabs $\rightarrow$  tubes $\rightarrow$ bubbles. The parameter sets in the latter category are the ones that seem to give a higher density ($\rho_c$) at which the crust-core transition takes place. As one can see that the appearance of different pasta structures is sensitive to the applied model,  one needs to investigate the model dependence.

The sensitivity of pasta phase appearance can be attributed to two main factors: a) the parametrization of surface and curvature energy and b) the EoS for the bulk and neutron gas surrounding the clusters. Since pasta phase appearance takes place in the region where matter is highly neutron-rich, the correct parametrization of surface and curvature tension Eqs. (\ref{surf} and \ref{curv}) becomes important. For this, we fit the surface and curvature parameters in Eqs. (\ref{surf}) and (\ref{curv}) i.e. parameter space $\boldsymbol{S}= \{\sigma_0, b_s, \sigma_{0,c}, \beta,\alpha,p\}$ with the experimental atomic mass evaluation of AME2020 \cite{Huang_2021} using a suitable penalty function \cite{Dobaczewski_2014, Parmar_2022, Carreau_2019}.  The surface energy plays a seminal role in determining the crustal properties of the neutron star, and therefore, fitting this parameter space for individual EoS is essential to appropriately estimate the surface energy rather than using the same value for all the models. Additionally, there exists a minor energy difference between various pasta structures \cite{Pearson_2018}, and hence, the finite size corrections in terms of surface and curvature term become crucial.   
The value of $p$, which takes care of the isospin asymmetry dependence of surface energy, is taken to be 3. This is a favorable choice in various calculations of surface energy \cite{Lattimer_1991, Avancini_2009, Carreau_2019, Avancini_2009}, A lower/higher value of the surface parameter $p$ results in a larger/smaller value of the surface tension. A smaller surface tension consequently predicts larger crust-core transition density and pressure (see Fig. 6 of \cite{Parmar_2022}). This further impacts the sequence of pasta configuration in Fig. \ref{fig:past_bar}. In our calculations, we vary the value of $p$ from $2.5$ to $3.5$ and observe that the number of pasta structures does not change, but the density at which they occur increases slightly for the higher value of $p$. We  take the value of  $\alpha$ to be 5.5  as per Ref. \cite{Newton_2013} . Values of rest of the parameter space $\boldsymbol{S}$ is given in Table \ref{tab:surfaceparameter} for all the models considered in Table \ref{tab:forceproperties}. It is evident that the surface parameter $b_s$ has the largest deviation among $ \{\sigma_0, b_s, \sigma_{0,c}, \beta\}$. The $b_s$  characterizes the change in the surface and curvature tensions for small deviations from isospin symmetry. Furthermore, our choice of the simplified mass formula, Eq. \ref{eq:energy}, is conceptually limited by the fact that mere knowledge of the nuclear mass is not sufficient to derive the surface and curvature contribution because of the partial compensation between nuclear bulk and the surface. Although we have not explicitly considered the shell energy contribution in the nuclear mass, they are bound to be implicitly accounted for in the fitting procedure by optimizing the values of surface parameter space $\boldsymbol{S}$. 
\begin{table}
\centering
\caption{The fitted value of surface and curvature energy parameters for various force parameters. The value of $\alpha$ and $p$ is taken to be 5.5 and 3, respectively. Experimental binding energy is taken from AME2020 table \cite{Huang_2021}. }
\label{tab:surfaceparameter}
\scalebox{1.1}{
\begin{tabular}{lllll}
\hline
\hline
Parameter & 
\begin{tabular}[c]{@{}l@{}} \hspace{0.2cm} $\sigma_0$ \\(MeV fm$^{-2}$) \end{tabular}& \hspace{0.2cm} $b_s$ &
\begin{tabular}[c]{@{}l@{}} \hspace{0.2cm} $\sigma_{0,c}$ \\(MeV fm$^{-1}$) \end{tabular} & \hspace{0.2cm} $\beta$ \\
\hline
BKA24 & 0.99339 & 14.3342 & 0.07965 & 0.7711 \\\hline
FSU2 & 0.96665 & 8.77776 & 0.09014 & 0.88746 \\\hline
FSUGarnet & 1.13964 & 29.3893 & 0.07844 & 0.44268 \\\hline
G1 & 0.93641 & 5.55101 & 0.09977 & 0.97866 \\\hline
G2 & 0.99538 & 8.81859 & 0.09672 & 0.85788 \\\hline
G3 & 0.88424 & 26.5837 & 0.09921 & 0.93635 \\\hline
GL97 & 0.73897 & 17.1523 & 0.12018 & 1.19306 \\\hline
IOPB-I & 0.97594 & 16.3546 & 0.09064 & 0.81485 \\\hline
IUFSU & 1.19953 & 30.2177 & 0.07691 & 0.31875 \\\hline
IUFSU* & 1.04205 & 34.2857 & 0.08197 & 0.62258 \\\hline
SINPA & 1.02767 & 24.5575 & 0.08667 & 0.69476 \\\hline
SINPB & 1.03574 & 15.2161 & 0.08332 & 0.70222 \\\hline
TM1 & 0.79998 & 7.35242 & 0.10278 & 1.14013\\\hline \hline
\end{tabular}%
}
\end{table}

\begin{figure}
    \centering
    \includegraphics[width=0.5\textwidth]{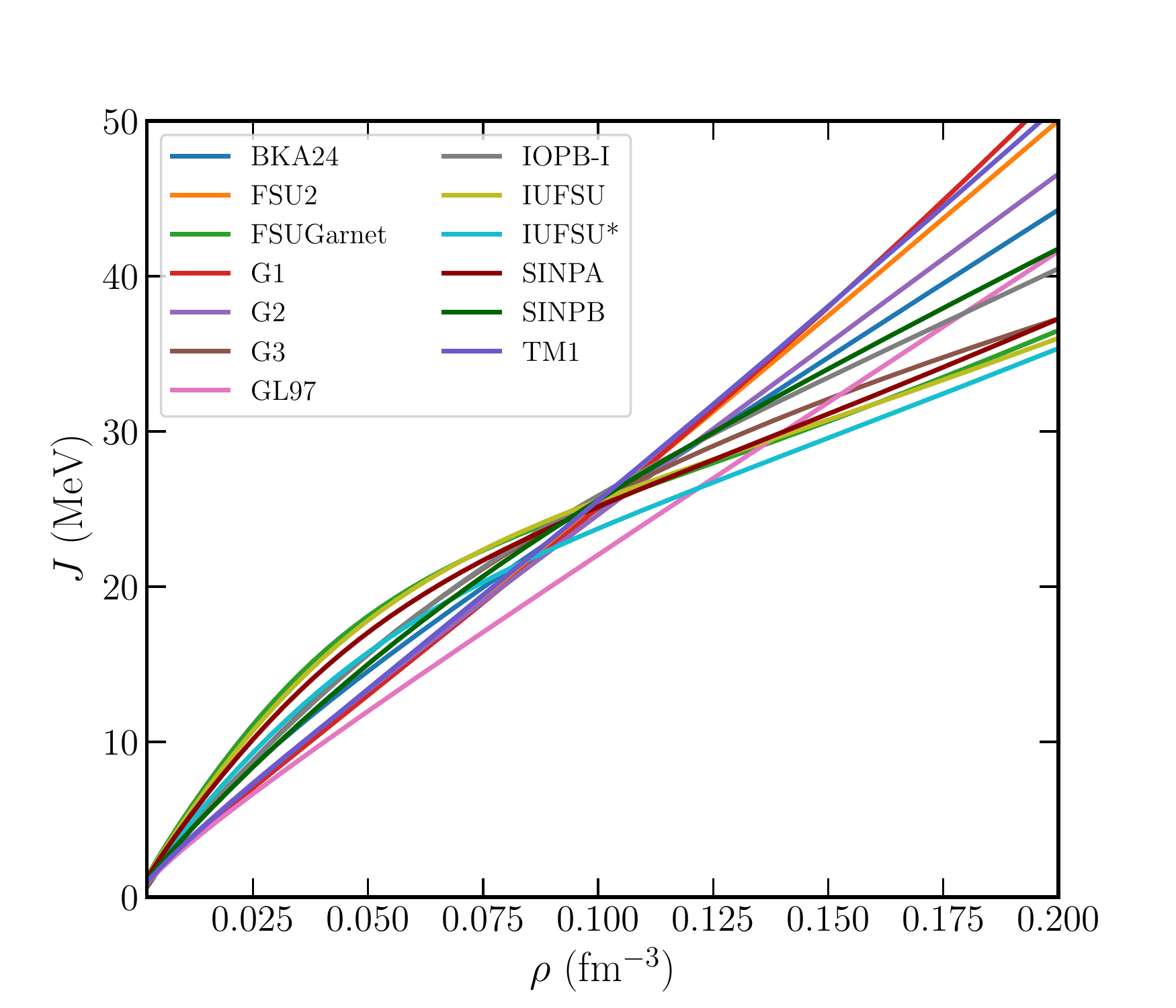}
    \caption{Symmetry energy of the models considered in Fig. \ref{fig:past_bar}.}
    \label{fig:symmetryenergy}
\end{figure}

\begin{figure*}
    \centering
    \includegraphics[width=0.75\textwidth]{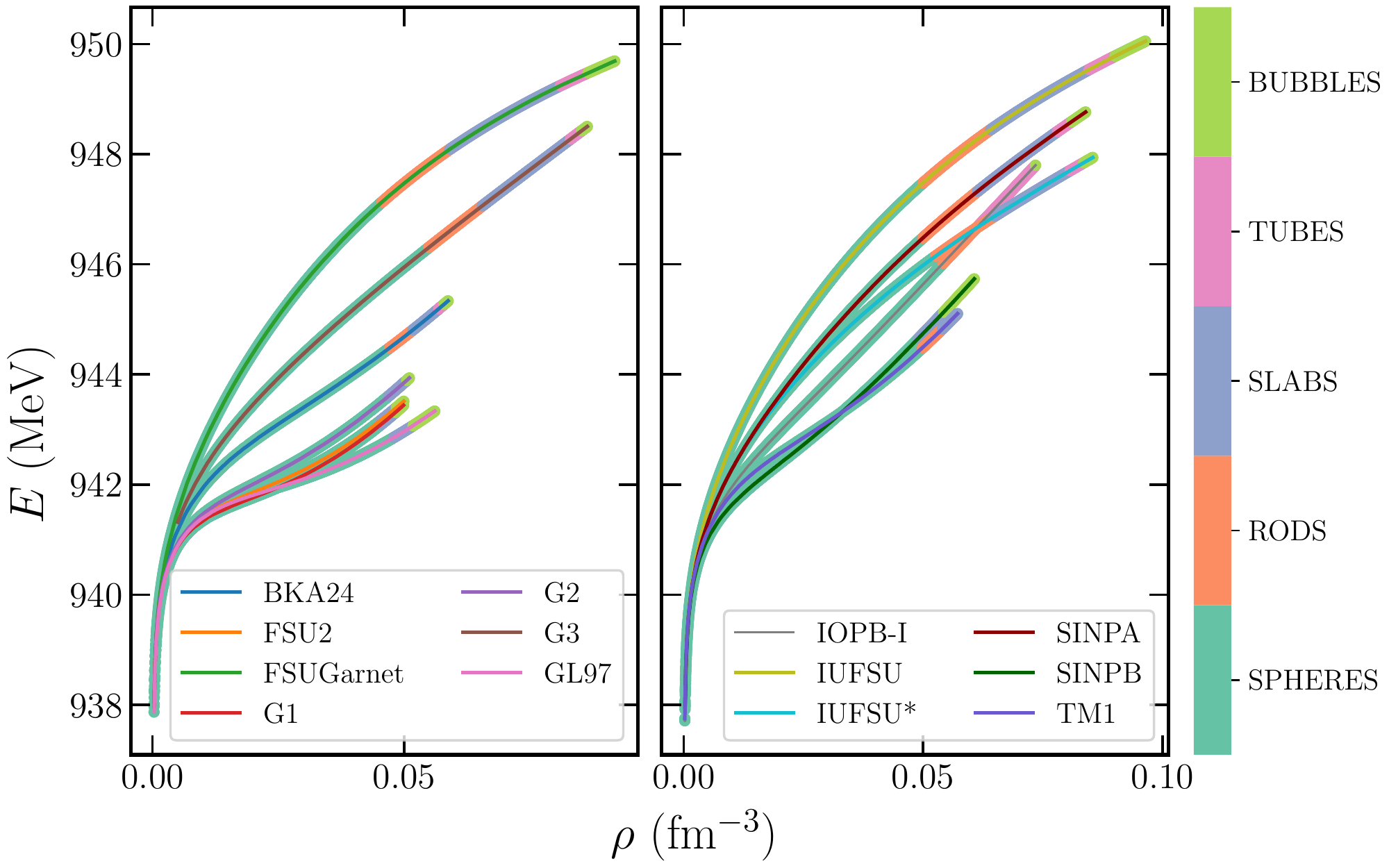}
    \caption{The equilibrium value of WS cell energy for various parameter sets considered in Fig. \ref{fig:past_bar} with the range of different pasta structures. }
    \label{fig:pasta_eos}
\end{figure*}

Calculation of the inner crust composition is a problem of two-phase equilibrium, which is solved using suitable mechanical and dynamical equations \cite{Parmar_2022, Carreau_2019, Vishal_2021}. In such a system, the symmetry energy plays a deciding role \cite{Vishal_2021, Alam_2017} and is known to influence the inner crust EoS \cite{Pearson_2018}. Furthermore, with ever-improving astrophysical data, establishing available correlations among various nuclear matter and neutron star observables is highly desirable to constrain the equation of state. Nuclear matter properties such as symmetry energy, slope parameter, etc., are calculated at saturation density. These correlations are crucial to fine-tune the theoretical models. Since the relevant density range for crust properties of neutron stars lies below subsaturation density, i.e., below 0.1 fm$^{-3}$, 
one should not merely compare the crust properties of neutron stars with the saturation value of nuclear matter observables. To access the role of symmetry energy on crust EoS, we show in Fig. \ref{fig:symmetryenergy} the density dependence of symmetry energy ($J$) for the parameter sets and the corresponding behavior of equilibrium value of WS cell energy of the inner crust in Fig. \ref{fig:pasta_eos}. 
The density dependence of symmetry energy in the subsaturation density region seems to impact the WS cell energy directly. The parameter sets such as FSUGarnet, G3, IOPB-I, IUFSU, and IUFSU* show higher symmetry energy in the subsaturation density and hence higher crust-core transition density. These forces predict all five pasta phases. The parameter set BKA24, however, estimates lower symmetry energy yet predicts all the five pasta phases. The remaining forces, which estimate lower symmetry energy, estimate the possibility of only three pasta phases, i.e., sphere, cylinder, and slab, and lesser WS energy as shown in Fig. \ref{fig:pasta_eos}. It is relevant to mention that the behavior of symmetry energy is different below and above the subsaturation density region, i.e., half the value of saturation density. Therefore, one must be cautious while analyzing the impact of symmetry energy on low-density EoS. We provide the values of $J$ and $L$ at saturation density and $\rho=0.05$fm$^{-3}$ in Table \ref{tab:forceproperties}. Since the relative behavior of symmetry energy among the considered force parameter somewhat remains the same below 0.075 fm$^{-3}$, therefore, the value of 0.05 fm$^{-3}$ is taken as a reference.
\begin{figure*}
    \centering
    \includegraphics[width=.7\textwidth]{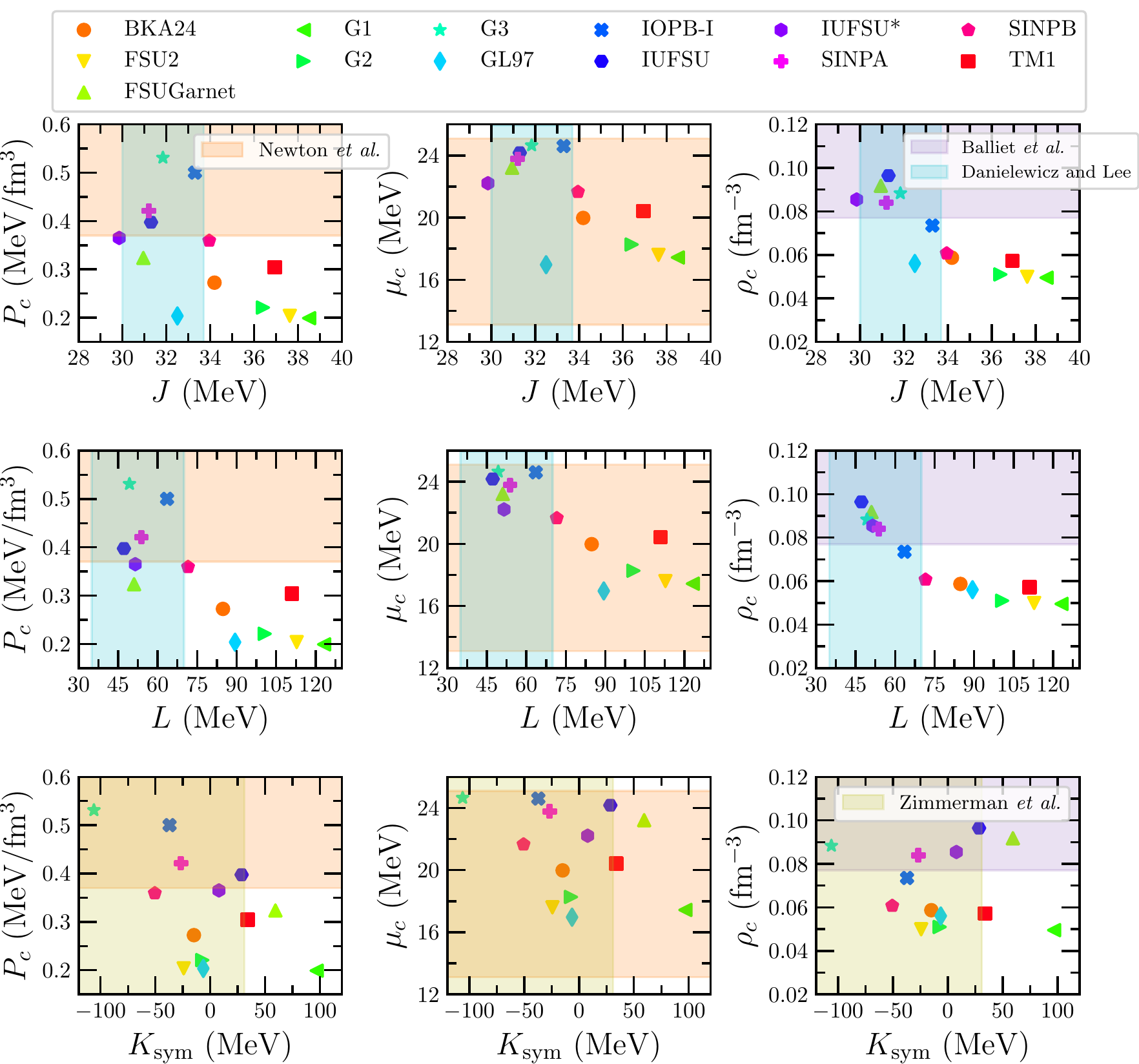}
    \caption{The crust-core transition pressure $P_c$, chemical potential $\mu_c$, and density $\rho_c$ as a function of symmetry energy $J$, slope parameter $L$ and $K_{\rm sym}$. The orange band represent the median range  obtained in Newton {\it et al.} \cite{Newton_2021} for the uniform Prior + PREX \cite{Adhikari_2021}  data while the purple band represent the uniform Prior + PNM band from the Balliet {\it et al.} \cite{Balliet_2021} for 95\% credible range. The vertical cyan band for the empirical/experimental range of symmetry energy and its slope parameter constraints given By Danielwicz {\it et al.} \cite{Danielewicz_2014}. The olive vertical band represents the $K_{\rm sym}$ constraints by Zimmerman {\it et al.} \cite{Zimmerman_2020}.}
    \label{fig:pcc}
\end{figure*}

\begin{figure*}
    \centering
    \includegraphics[width=0.7\textwidth]{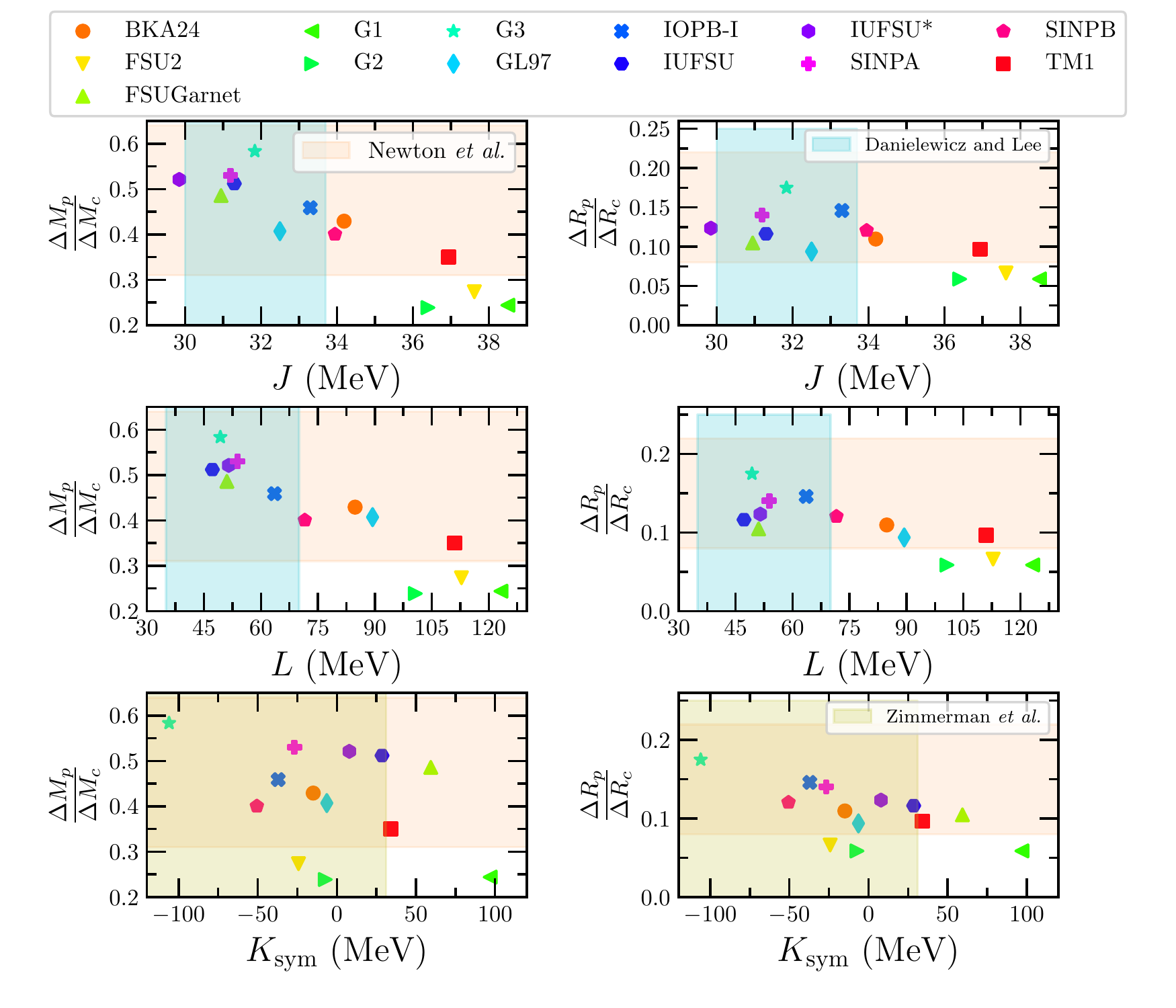}
    \caption{The mass/moment of inertia and thickness fractions of pasta as a function of symmetry energy $J$, slope parameter $L$ and $K_{\rm sym}$}
    \label{fig:pasta_mass_thick}
\end{figure*}
As the density grows in the inner crust, the clusters' surface tension increases, and the system favors the homogeneous phase energetically. We calculate the transition from the heterogeneous crust to a homogeneous core where the energy of the WS cell becomes equal to the energy of the core,  $E_{\rm WS}(\rho_c)=E_{npe\mu}(\rho_c)$. However, it is not the transition density ($\rho_c$) that determines the location of the crust–core boundary but the transition pressure and chemical potential \cite{Balliet_2021}. The transition pressure controls the mass and moment of inertia of the crust (see Eqs. \ref{eq:pp} and \ref{eq:moic} ) while the transition chemical potential determines the thickness of the crust and the pasta structures. In Fig. \ref{fig:pcc} we compare the transition pressure $P_c$, chemical potential $\mu_c$ and density $\rho_c$ as a function of symmetry energy $J$ and its higher order derivatives, slope parameter $L$ and curvature $K_{\rm sym}$ at the saturation density for various forces,  with the constraints obtained from the Bayesian inference analysis from the two separate studies of Newton \textit{et al.} \cite{Newton_2021} and Balliet \textit{et al.} \cite{Balliet_2021} which use  an extended Skyrme energy density functional within CLDM.

The E-RMF models that satisfy the Newton {\it et al.} prior + PREX data are the ones that have a lower value of $J$ and $L$ in accordance with the isobaric analog states data \cite{Danielewicz_2014}. However, only  parameter sets SINPA, FSUGarnet, IUFSU, IUFSU$^*$ and TM1 satisfy a more stringent constraints on $P_c$  based on  Skins+ PNM data 
which  results in $P_c=0.38^{+0.08}_{-0.09}$. In contrast, all these model satisfy the prior + PNM constraint of Balliet {\it et al.} \cite{Balliet_2021} which predict it to be $P_c= 0.49^{+0.27}_{-0.28}$ MeV fm$^{-3}$ on 95\% credible range. All the parameter sets estimate the transition chemical potential $\mu_c$  in agreement with the Newton \textit{et al.} \cite{Newton_2021}. At the same time, the models with lower symmetry energy do not obey the range of $\mu_c=14.7^{+4.7}_{-5.0}$ given by Balliet {\it et al.} \cite{Balliet_2021}. For the transition density, only models IUFSU, IUFSU*, SINPA, G3, and FSUGarnet satisfy the available constraint from  Balliet {\it et al.} \cite{Balliet_2021}. Furthermore, $P_c$, $\mu_c$, and $\rho_c$ seem to decrease with higher values of $J$, $L$ and $K_{\rm sym}$ advocating the role of symmetry energy on the crust parameters. The relationship of $K_{\rm sym}$ with $P_c$, $\mu_c$ and $\rho_c$ appears to have a large variance compared to the $J$ and $L$. It should be mentioned here that the transition density is almost half the value of saturation density where the respective values of $J$, $L$, and $K_{\rm sym}$ are calculated. Therefore, the above relationships should accompany the knowledge of symmetry energy in the subsaturation region \cite{Balliet_2021}.
\subsection{Relative pasta layer thickness and mass}
\label{result:relativepasta}
Various theoretical calculations predict that the pasta structures account for more than 50\% of the mass of the crust and 15\% of its thickness \cite{Newton_2021, Balliet_2021, Dinh_2021, Fabrizio_2014}. In view of this, the following Ref.
\cite{Lattimer_2007}, we calculate the mass and thickness of the nonspherical shapes using the E-RMF models and compare them with the available theoretical range. The main ingredients are the chemical potential and pressure defined in section \ref{relativepasta}. In Fig. \ref{fig:pasta_mass_thick}, we show the relative mass and the thickness of the nonspherical shapes as a function of $J$, $L$, and $K_{\rm sym}$. All the models except G1, G2, and FSU2, which estimate a relatively larger value of symmetry energy and slope parameter, predict the nonspherical pasta mass and thickness within the range calculated by Newton {\it et al.} from PREX constraints. These are also consistent with the Skins+ PNM constraints of the  Newton {\it et al.} \cite{Newton_2021} ( $\frac{\Delta M_p}{\Delta M_c}= 0.49^{+0.06}_{-0.11}$ , $\frac{\Delta R_p}{\Delta R_c}= 0.132^{+0.023}_{-0.041}$), posterior estimations of Thi {\it et al.} \cite{Dinh_2021} ( $\frac{\Delta M_p}{\Delta M_c}= 0.485 \pm 0.138$, $\frac{\Delta R_p}{\Delta R_c}= 0.128 \pm 0.047$) using meta-model formalism \cite{Carreau_2019} and with the prior + PNM range of Balliet {\it et al.} \cite{Balliet_2021} ( $\frac{\Delta M_p}{\Delta M_c}= 0.62 ^{+0.03}_{-0.04}$  and $\frac{\Delta R_p}{\Delta R_c}= 0.29^{+0.04}_{-0.09}$).  Since the mass fraction is directly proportional to the amount of moment of inertia \cite{Lorenz_1993}, the behavior of pasta mass also holds good for its moment of inertia content. Furthermore, the parameter sets with smaller $J$, $L$, and $K_{\rm sym}$  seem to give a larger mass and thickness of the pasta structure. A linear relationship between mass and thickness of pasta with $J$, $L$, and $K_{\rm sym}$ is also evident. 
\begin{figure}
    \centering
    \includegraphics[width=0.45\textwidth]{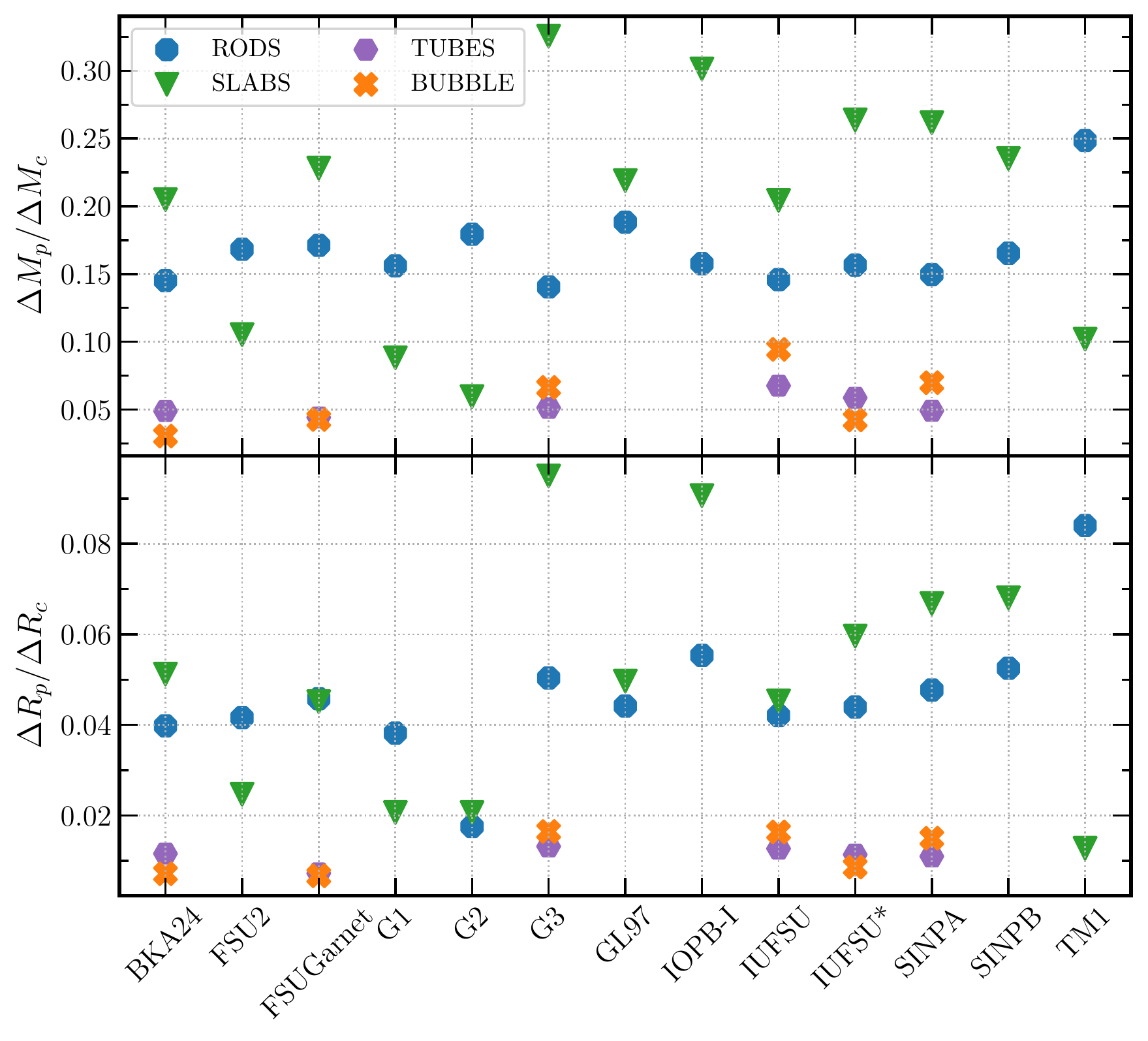}
    \caption{ Upper panel shows the relative mass of the different layers of pasta structures, and the lower panel shows the relative thickness compared to the total crust.}
    \label{fig:rel_mass_thick}
\end{figure}

In Fig. \ref{fig:rel_mass_thick} we show relative mass and thickness of different layer of pasta in the inner crust using the same method as for the total pasta content (Eqs.  \ref{eq:rr} and \ref{eq:pp}). In our calculations of pasta phases, we see that all the models at least predict two nonspherical phases, namely, rods and slabs. The rod pasta phase has mass $\approx$ 15\% of the mass of the crust except for the TM1 set, which estimates its mass $\approx$ 25\%. The thickness of this phase is $\approx$ 4\% of the crust thickness. The parameter sets that predict the existence of only two nonspherical pasta phases before transiting into the homogeneous core have the mass and thickness of the slab phase lesser than the rod phase. The IOPB-I has an exception among these sets. It may be noted that we do get a third nonspherical tube phase for the IOPB-I set but within a small density range, and hence we do not consider the rod phase for IOPB-I (see Fig. \ref{fig:past_bar}). Once again, the symmetry energy seems to impact the relative amount of pasta structures. The parameter sets such as TM1, FSU2, G1, and G2 that have lower symmetry energy in the subsaturation density region predict the larger contribution of the rod phase compared to the slab phase. The remaining parameter sets predict the largest mass and thickness fraction for the slab phase. It accounts for $\approx$ 20\% of the crust mass and 5\% of the crust thickness. The G3 and IOPB-I sets estimate them as large as 30\% and 9\%, respectively. The tube and bubble phase has the smallest content in the inner crust. They account for about 5\% of the crust mass and 1\% of the thickness, subject to their occurrence.  

It is apparent that the existence of pasta structures in the inner crust is greatly influenced by the nuclear EoS. The density dependence of symmetry energy has a prominent role in determining their mass and thickness. To quantify the relationships discussed above, we carry out a Pearson correlation analysis of various crust properties.
\begin{figure}
    \includegraphics[width=0.5\textwidth]{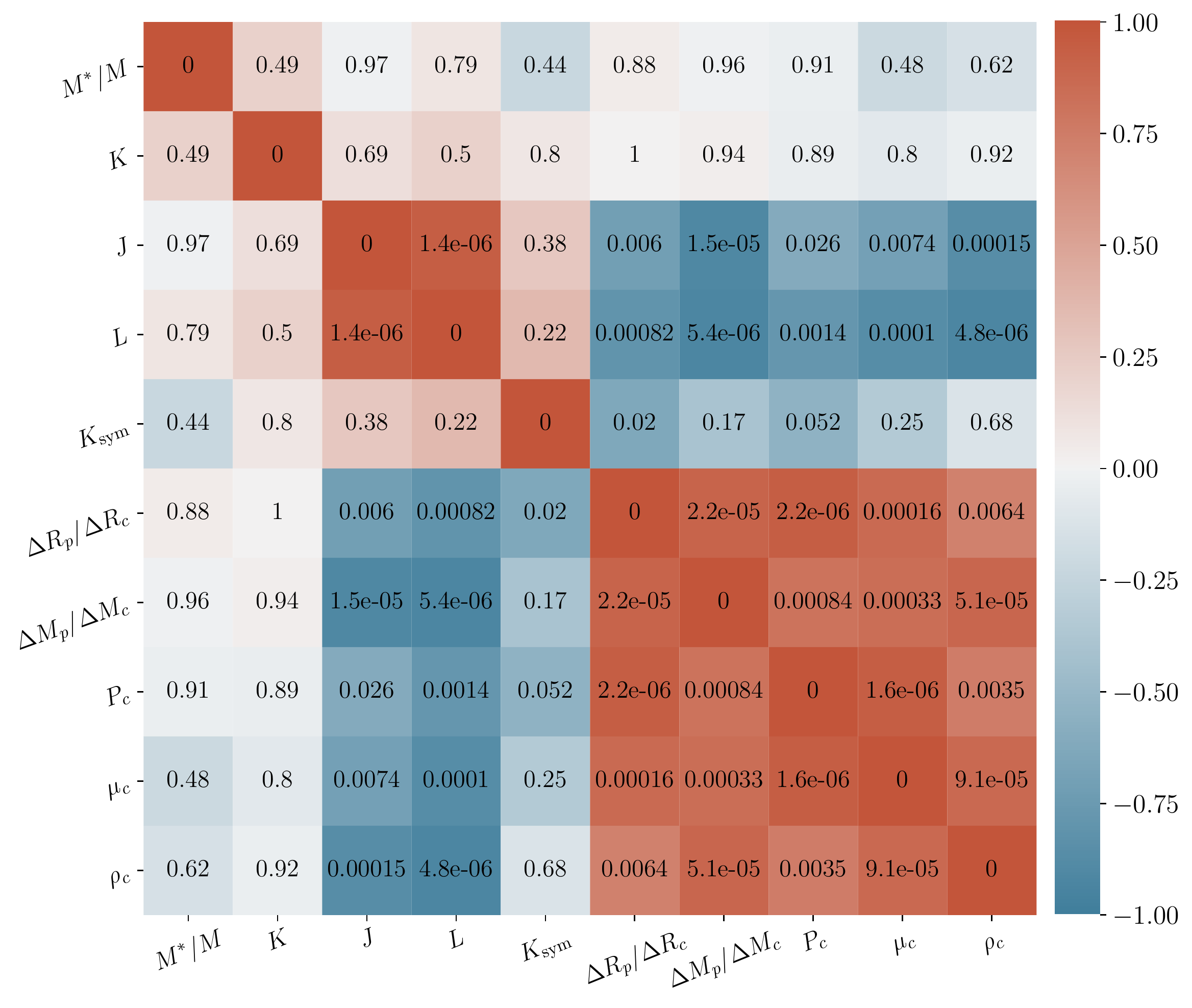}
    \caption{Correlation heat map of the bulk properties with the crustal properties and surface parameters. The color map signifies the strength of the correlation while the values represent the associated $p$-values.}
    \label{fig:corr}
\end{figure}
Fig. \ref{fig:corr} shows the Pearson correlation matrix between the bulk properties, effective mass ($M^*/M$), incompressibility ($K$), symmetry energy ($J$), slope parameter ($L$) and curvature of symmetry energy ($K_{\rm sym}$) with crustal properties namely relative thickness ($\frac{\Delta R_p}{\Delta R_c}$) and mass of the pasta ($\frac{\Delta M_p}{\Delta M_c}$) along with the transition pressure ($P_c$), chemical potential ($\mu_c$) and density ($\rho_c$). The color shows the strength of the correlation while the values represent the statistical significance in the form of $p$-value or probability value \cite{LAN1959testing}. A $p$-value signifies the statistical significance of the used statistics (here Pearson correlation), and a value less than 0.05/0.01 is generally considered statistically significant for a 95/99\% interval. It is seen that the bulk properties $M^*/M$ and $K$ do not correlate with the crustal properties. On the other hand, symmetry energy and slope parameter show a strong negative correlation with pasta mass and thickness along with the transition pressure, chemical potential, and density within a 95\% confidence interval. These relations are consistent with those obtained in previous studies \cite{Oyamatsu_2007, Newton_2012}. The $K_{\rm sym}$ shows some negative correlation with the relative thickness of the pasta. 

Additionally, the pasta's mass and thickness are strongly correlated with the transition pressure, chemical potential, and density. All of these relations, which are obtained within the E-RMF framework along with the CLDM formalism, are consistent with the recent work based on Bayesian inference of the neutron star crust \cite{Newton_2012, Balliet_2021, Dinh_2021}. Although these works are based on the relatively more straightforward nuclear interaction models as per the requirement of Bayesian analysis, they provide us with the relevant estimation of various crust properties. The E-RMF model considered in this work is all within reasonable agreement with the theoretical constraints and therefore suitable for further structural calculations of numerous neutron star properties such as superfluidity, conductivity, etc.    
\subsection{Shear modulus and torsional oscillation mode}
A magnetar, which is an exotic type of neutron star, is characterized by an extremely high magnetic field of the order of $10^{15}$G, which results in the powerful x-ray emission powered by the reconfiguration of the decaying field. The rapidly evolving field, when it strikes the solid crust, results in an associated starquake, detectable as quasiperiodic oscillations (QPOs) \cite{Steiner_2009, Thompson_2001} in the tails of light curves of giant flares from soft gamma-ray repeaters (SGR) \cite{Israel_2005, Strohmayer_2005}. In this context, it becomes essential to understand the shear property of the crust. The shear modulus, which describes the crust's elastic response under the shear stress, leads to the shear oscillations. The shear oscillations travel through the star's crust with shear velocity ($V_s$). These shear modulus and shear velocity are the characteristics of the crust composition, which consequently depends on the nuclear EoS and surface energy parametrization. We use the Monte Carlo simulation results in the form of Eq. (\ref{eq:shearmodulus}) for the spherical portion of the inner crust. The elastic response of the nonspherical phase is not yet fully understood, but the crust's rigidity is expected to decrease and vanish at the crust-core boundary \cite{Gearheart_2011,Passamonti_2016}. To model the shear modulus in this region, we use Eq. (\ref{eq:mubar}). 
\begin{figure}
    \centering
    \includegraphics[width=0.5\textwidth]{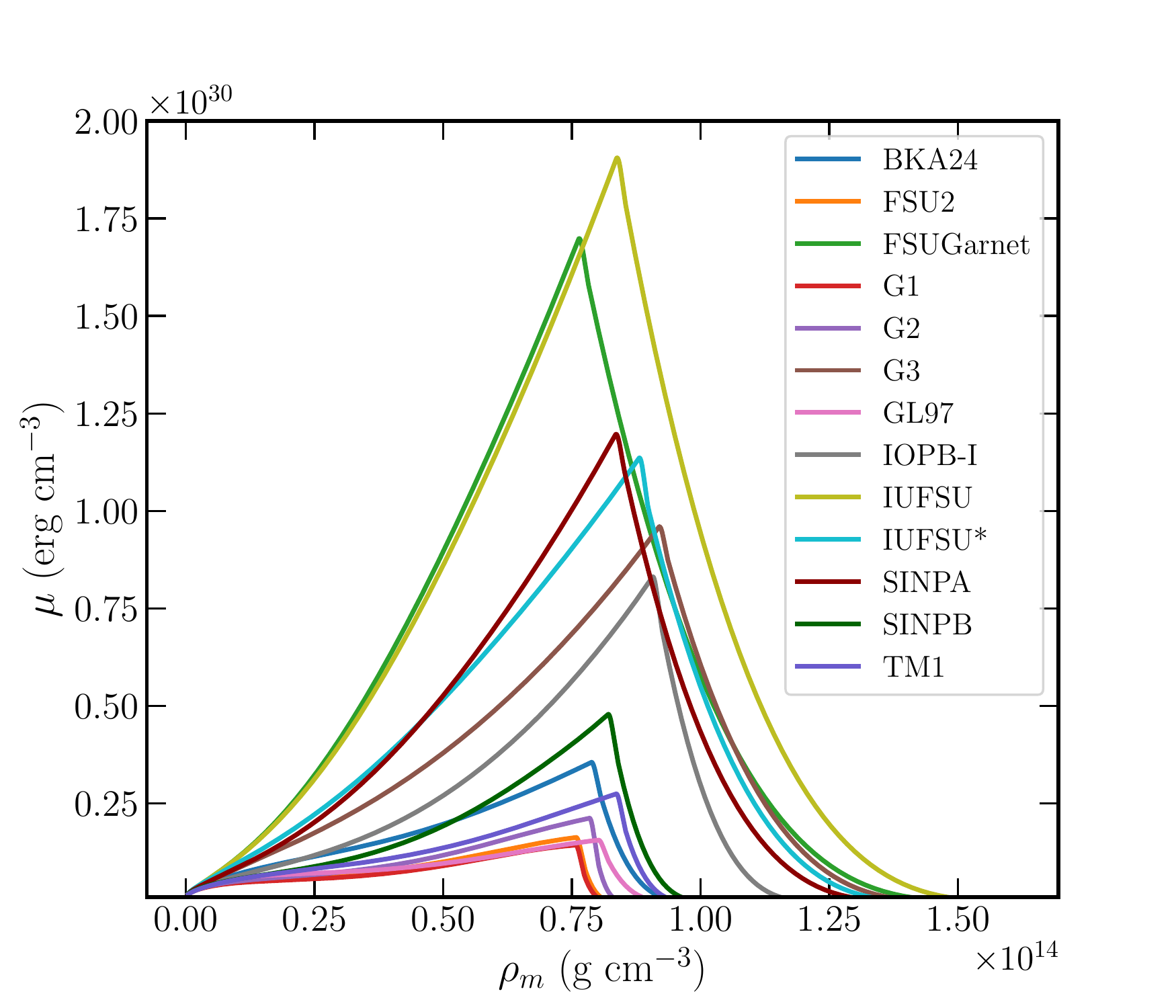}
    \caption{Shear modulus ($\mu$) of the inner crust for various E-RMF sets.}
    \label{fig:shear}
\end{figure}

The complete behavior of the shear modulus of the inner crust is shown in Fig. \ref{fig:shear}. As one moves deeper into the crust, the shear modulus increases monotonically until one reaches the density where the pasta phase appears. It then starts decreasing smoothly until the crust-core boundary and then vanishes. This behavior directly results from our approximation of the shear modulus in the pasta phase region. There is a significant uncertainty among different models, which is the consequence of the inner crust composition predicted by these models. Since the density dependence of symmetry energy and slope parameter predominantly control the inner crust \cite{Parmar_2022}, we see its effect on the shear modulus as well. In the subsaturation region, forces such as IUFSU, G3, BKA24, and FSUGarnet, which have a higher value of symmetry energy, estimate a larger shear stress value. 
\begin{figure}
    \centering
    \includegraphics[width=0.5\textwidth]{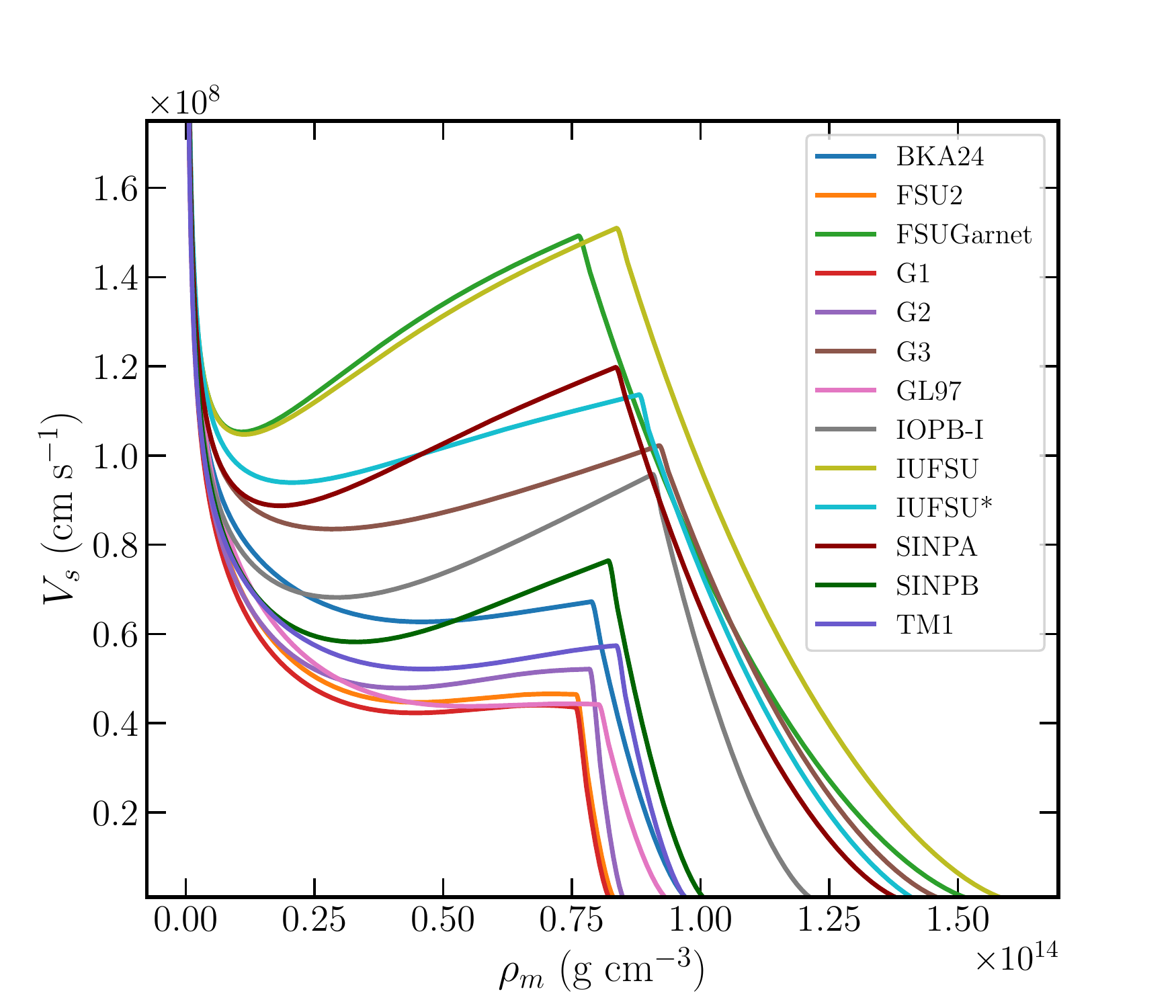}
    \caption{The shear velocity ($V_s$) as a function of the mass density for the various E-RMF models.}
    \label{fig:shear_speed}
\end{figure}

The shear velocity in the crust directly follows from the shear modulus and can be calculated using Eq. (\ref{eq:shearspeed}). In principle, the neutron superfluidity in the neutron star crust plays a crucial role in its dynamics \cite{Haskell_2018}. The superfluid neutrons are unlocked from the lattice moment and do not influence the shear modulus. However, Chamel  \cite{Chamel_2012} found that $\approx$ 90\% of the superfluid neutrons can be entrained to the lattice due to the Bragg scattering. In this work, however, we consider the dynamical mass in Eq. (\ref{eq:shearspeed}) equal to its total mass density \cite{Steiner_2009} neglecting the effect of the superfluidity and entrainment effects. The calculated shear speed will then underestimate its value, but the qualitative nature will remain unaffected. We show the behavior of shear speed in Fig. \ref{fig:shear_speed} for the corresponding shear modulus in Fig. \ref{fig:shear}. We do not show the shear speed in the outer crust, which is well established \cite{tews_2017} and increases with the increase in density. However, in the inner crust, it drops initially and then increases until the onset of the pasta structures. It decreases smoothly afterward and vanishes at the crust-core boundary. One can see that there is $\approx$ 3 fold difference between the lowest and highest value of shear speed among the parameter sets considered in this work. The dependence of shear velocity on the density also varies in a different way indicating the role of crust composition. 
\begin{figure}
    \centering
    \includegraphics[width=0.45\textwidth]{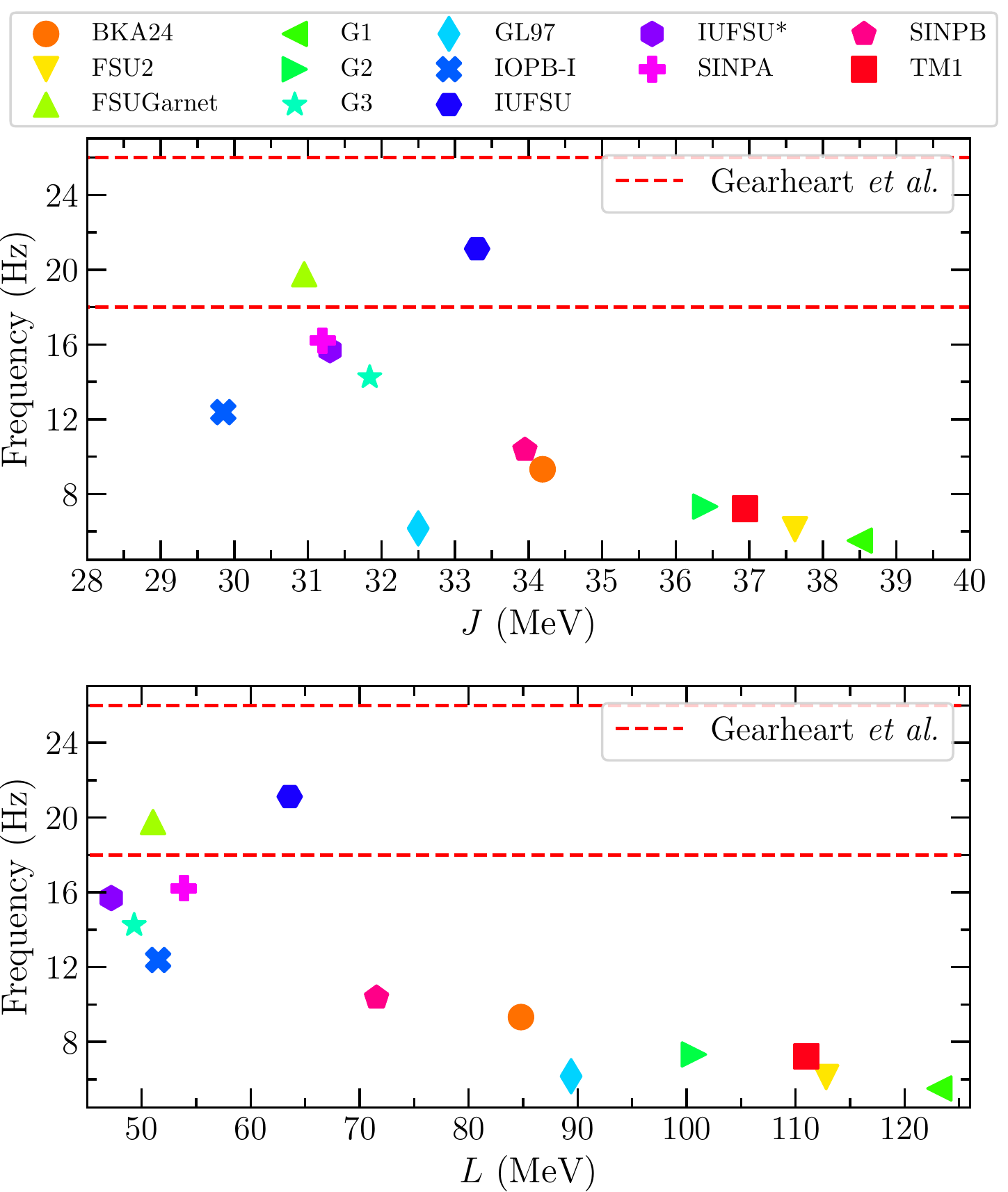}
    \caption{Frequency of fundamental torsional oscillation mode ($l=2$) in the crust for the maximum mass with $J$ and $L$. The two horizontal lines correspond to the observed value of 18 and 26 Hz. }
    \label{fig:freq}
\end{figure}

To approximately infer the fundamental torsional oscillation mode, we set the pasta shear modulus to be zero, considering the pasta as a liquid \cite{Gearheart_2011}. This means that the shear modulus and shear velocity in the solution of crustal shear perturbation equations (Eq. \ref{eq:freq}) are calculated at the boundary between the phase of spherical nuclei and the pasta phases, i.e., $\rho=\rho_{ph}$. We show the calculated frequencies of fundamental oscillation mode ($l=2$) for the maximum mass as a function of $J$ and $L$ for the various E-RMF models in Fig. \ref{fig:freq} along with the possible candidate of frequencies for the fundamental modes of QPOs: 18 Hz and 26 Hz \cite{Gearheart_2011, Israel_2005}. The fundamental frequency decreases with the symmetry energy and the slope parameter, which is consistent with the Refs. \cite{Gearheart_2011, Sotani_2013}. We see that considering the pasta phase to be liquid and ignoring superfluid and entrainment effects, the fundamental mode frequency agrees with the observed QPOs from SGRs at low symmetry energy and slope parameter. Only FSUGarnet and IUFSU parameter set match with the 18 Hz observed frequency. The frequencies also do not match with higher possible candidate frequencies of 28 Hz and  30 Hz \cite{Greif_2020, Israel_2005}. It may be noted that considering pasta to be liquid and neglecting entrainment effects reduces the frequency by a factor of $\approx$ 3  \cite{Gearheart_2011}. Therefore, the frequencies calculated in this work make the lower bound of the fundamental frequency. 
It is also clear that the pasta structures play a significant role in a crustal torsional mode. Moreover, the frequency modes in QPOs can be used as one of the asteroseismological sources to constrain the amount of pasta along with the nuclear matter observables such as symmetry energy and slope parameter, etc.  
\subsection{Neutron star observables}
\begin{table*}
\centering
\caption{The neutron star properties such as maximum mass ($M_{\rm max}$), radius corresponding to the maximum mass  ($R_{max}$), canonical radius ($R_{1.4}$), normalized maximum MI ($I_{max}$), normalized canonical MI ($I_{1.4}$), mass of the crust ($M_{\rm crust}$), thickness of the crust ($l_{crust}$), second Love number and dimensionless tidal deformability for canonical and maximum mass for 13 considered EoSs.}
\label{tab:NS_properties}
\renewcommand{\tabcolsep}{0.26cm}
\renewcommand{\arraystretch}{1.6}
\begin{tabular}{lllllllllllll}
\hline \hline
\begin{tabular}[c]{@{}l@{}}Parameter\\ sets\end{tabular} &
\begin{tabular}[c]{@{}l@{}}$M_{\rm max}$\\ ($M_\odot$)\end{tabular} &
\begin{tabular}[c]{@{}l@{}}$R_{\rm max}$\\  (km)\end{tabular} &
\begin{tabular}[c]{@{}l@{}}$R_{1.4}$\\ (km)\end{tabular} &
$I_{\rm max}$ &
$I_{1.4}$ &
$I_{\rm crust}/I$ &
\begin{tabular}[c]{@{}l@{}}$M_{\rm crust}$\\ ($M_\odot$)\end{tabular} &
\begin{tabular}[c]{@{}l@{}}$l_{\rm crust}$\\ (km)\end{tabular} &
$k_{2,1.4}$&
$\Lambda_{1.4}$ &
$k_{2,{\rm max}}$&
$\Lambda_{\rm max}$\\ \hline
BKA24&1.963 &11.61 &13.42 &0.401 &0.339 &0.0100  &0.008 &0.455 &0.0888 &681.25 &0.0307 &21.02
\\ \hline
FSU2&2.071 &12.12 &14.02 &0.405 &0.335 &0.0087 &0.008 &0.418 &0.0943 &899.64 &0.0302 &19.54
\\ \hline
FSUGarnet&2.065 &11.79 &13.19 &0.418 &0.343 &0.0100 &0.009 &0.542 &0.0889 &629.13 &0.0306 &17.54 \\ \hline
G1&2.159 &12.30 &14.15 &0.415 &0.331 &0.0084 &0.008 &0.413 &0.0922 &922.41 &0.0287 &16.34
\\ \hline
G2&1.937 &11.17 &13.27 &0.403 &0.333 &0.0069 &0.006 &0.378 &0.0819 &593.53 &0.0266 &16.11
\\ \hline
G3& 1.996 &10.95 &12.63 &0.425 &0.345 &0.0129 &0.011 &0.479 &0.0813 & 460.42 &0.0254 &11.98
\\ \hline
GL97&2.002 &10.81 &13.10 &0.423 &0.343 &0.0048  &0.004 &0.296 &0.0876 &596.43 &0.0221 &09.64
\\ \hline
IOPB-I&2.148 &11.96 &13.33 &0.428 &0.344 &0.0147 &0.014 &0.507 &0.0925 &686.49 &0.0292 &14.75
\\ \hline
IUFSU&1.939 &11.23 &12.61 &0.414 &0.351 &0.0110 &0.009 &0.510 &0.0871 &489.20 &0.0313 &19.10
\\ \hline
IUFSU*&1.959 &11.45 &12.92  &0.409 &0.347 &0.0114 &0.010 &0.526 &0.0880 &563.08 &0.0319 &20.66 \\ \hline
SINPA&2.000 &11.55 &12.93  &0.416  &0.349 &0.0136  &0.012 &0.515 &0.0908 &580.39 &0.0318 &19.36 \\ \hline
SINPB&1.993 &11.62 &13.16 &0.409 &0.342  &0.0128  &0.011  &0.486 &0.0881 &612.94 &0.0313 &19.98  \\ \hline
TM1&2.175 &12.36 &14.31 &0.415 &0.335 &0.0101 &0.009 & 0.444 &0.0979 &1037.5 &0.0285 & 15.97
\\ \hline \hline
\end{tabular}
\end{table*}
We model a complete neutron star by calculating the core EoS under the condition of charge neutrality \cite{Parmar_2022}, and $\beta$-equilibrium for each parameter set in Table \ref{tab:forceproperties} and make unified EoS by combining it with the inner crust EoS using the same parameter set along with the outer crust EoS discussed in Sec. \ref{formulation}. The unified EoSs are available  publicly in GitHub page\footnote{\url{https://github.com/hcdas/Unfied_pasta_eos}}. The unified treatment of each EoS ensures that the neutron star properties such as crust mass, thickness, the moment of inertia, etc., can be estimated and analyzed quite precisely. To calculate the neutron star observables, we solve the TOV Eqs. \ref{eq:pr} and \ref{eq:mr} for a fixed central density to obtain the $M-R$ profile, second Love number, and dimensionless tidal deformability. The moment of inertia is calculated under the slow rotation approximation using Eq. \ref{eq:moi}. We determine the total crust mass and thickness by integrating the TOV Eqs. \ref{eq:pr} and \ref{eq:mr} from $R=0$ to $R=R_{\rm core}$, which depends on pressure as $P(R=R_{\rm core})=P_t$. Finally, the crustal moment of inertia is worked out using Eq. \ref{eq:moic}. The detailed formalism of these quantities is provided in Refs. \cite{Lattimer_2000, Parmar_2022}. The mass and thickness of the crust for IOPB-I EoS are 0.013 $M_\odot$, and 0.490 km, respectively, without considering the pasta phase inside the crust (see Table 7 in Ref. \cite{Parmar_2022}). However, they are estimated to be 0.014 $M_\odot$ and 0.507 km, respectively, including the pasta structures. Hence, we notice that the crustal mass doesn't change, but the crustal thickness increases slightly when we consider pasta phases inside the crust.

We give the tabulated data for neutron star properties such as maximum mass ($M_{\rm max}$), radius corresponding to the maximum mass ($R_{\rm max}$), canonical radius ($R_{1.4}$), normalized maximum MI ($I_{\rm max}$), normalized canonical MI ($I_{1.4}$), mass of the crust ($M_{\rm crust}$), thickness of the crust ($l_{\rm crust}$), second Love number ($k_2$) and dimensionless tidal deformability ($\Lambda$)  for canonical and maximum mass for 13 considered EoSs in Table \ref{tab:NS_properties}. The maximum mass of all the sets reasonably satisfy the observational constraint  of massive pulsars such as PSR J0348+0432 ($M = 2.01\pm{0.04} \ M_\odot$) \cite{Antoniadis_2013} and PSR J0740+6620 ($M = 2.14_{-0.09}^{+0.10} \ M_\odot$) \cite{Cromartie_2019}. They are also  in accordance with the  radius constraints given by Miller {\it et al.} \cite{Miller_2019}, Riley {\it et al.} \cite{Riley_2019} and  PSR J0030+0451 with X-ray Multi-Mirror Newton for canonical star with $R_{1.4} = 12.35 \pm 0.75$ km \cite{Miller_2021}.

The normalized moment of inertia for slowly rotating NS is calculated for 13 EoSs. The numerical values are given in Table \ref{tab:NS_properties} both for the canonical and maximum mass star. There exists a Universal relation between the MI and the compactness of the star \cite{Lattimer_2005, Steiner_2016, LATTIMER_2016}. We compare the numerical values of $I$ with and without pasta phases as done in our earlier work \cite{Parmar_2022}. The value of $I_{\rm max}$ and $I_{1.4}$ for IOPB-I EoS was found to be 0.429 and 0.346 respectively (see Table 7 in Ref. \cite{Parmar_2022}) without pasta phases. By including the pasta phase, the values are slightly lesser and found to be 0.428 and 0.344, respectively. Similar cases are seen both for FSUGarnet and G3 EoSs. Hence, we observe that the pasta phases don't significantly influence the moment of inertia of the star. However, the crustal moment of inertia ($I_{\rm crust}/I$) for maximum mass estimated from these EoSs are consistent with the fractional moment of inertia (FMI) observed from the 581 pulsar glitches catalog \cite{Espionza_2011, Parmar_2022}. One can also see that the mass of the crust ($M_{\rm crust}$) is equivalent to the crustal moment of inertia, advocating the importance of unified treatment of crust and core equation of state. 

The Love number and dimensionless tidal deformability for only quadrupole case ($l=2$) are calculated as described in Ref. \cite{Das_2022}. The numerical values are given in Table \ref{tab:NS_properties} for  considered EoSs. For a realistic star, the value of $k_2$ is 0.05--0.1 \cite{Hinderer_2008}. Our calculated results are well within this range. The constraint on $\Lambda_{1.4}$ given by LIGO/Virgo \cite{Abbott_2017, Abbott_2018} from the binary neutron star merger event GW170817 with, $\Lambda_{1.4}=190_{-70}^{+390}$. Only G3, IUFSU, and IUFSU* are within the GW170817 limit. We also observed that the effects of pasta on both $k_2$ and $\Lambda$ are not significant as compared with only the spherical shape considered inside the crust. 

The relativistic nuclear models considered in this work suggest that $\approx$ 50\% of the crust mass and $\approx$ 15\% of the crust thickness is contained in the pasta structures. 
Since the entire crust itself comprises only 0.5-1\% of the neutron star mass and 5-10 \% of the radius, the pasta structures do not significantly impact the global properties of a neutron star such as maximum mass, the moment of inertia, Love number, dimensionless tidal deformability, etc. However, the pasta structure affects the microscopic properties of the neutron star, which essentially depend on the crust structure. The shear modulus, which determines the torsional oscillation mode of quasiperiodic oscillations (QPOs), is greatly influenced by the presence of pasta structures. The fractional crustal moment of inertia or mass is an important property to explain the pulsar glitches. The pasta content in the crust influences these properties by controlling the surface thickness. These structures also influence the magnetic field's decay rate, which explains the observed population of isolated X-ray pulsars \cite{Caplan_2017} and limits the maximum spin period of rotating neutron stars \cite{Pons_2013}. The properties such as viscosity, conductivity, neutrino cooling, etc., are also influenced by the nature of the structure present in the inner crust \cite{Caplan_2017}.
\section{Conclusion}
\label{conclusion}
In summary, we investigate the existence of pasta structures in the inner crust of a neutron star employing the compressible liquid drop model along with the effective relativistic mean-field theory. We consider three geometries: spherical, cylindrical, and planar, resulting in five configurations, namely sphere, rod, slab, tube, and bubble. The equilibrium configuration at a given baryon density is obtained by minimizing the energy of the five pasta structures. The main ingredient in calculating the inner crust is the proper treatment of the surface energy parametrization. In view of this, we optimize the surface and curvature tension based on Thomas-Fermi calculations for a given equation of state on recent atomic mass evaluation \cite{Huang_2021}.
   
In our calculations, we have used 13 well-known parameter sets that satisfy the recent observational constraints on the maximum mass and radius of the neutron star. We construct unified EoS for each of these sets to obtain the pasta and crustal properties consistently. The appearance of different pasta layers is model-dependent. The model dependency is attributed to the behavior of symmetry energy in the subsaturation density region and the surface energy parametrization. A thicker crust favors the existence of more number of pasta layers in it. We calculate the pressure ($P_c$), chemical potential ($\mu_c$), and density ($\rho_c$) of the crust-core transition from the crust side and compare the results with recent constraints proposed using Bayesian inference analyses \cite{Newton_2021, Balliet_2021}. The parameter set with lower values of $J$, $L$, and $K_{\rm sym}$ seem to agree better with these theoretical constraints. 

It is seen that the ($P_c$) and ($\mu_c$) play a more critical role in determining the crust structure instead of ($\rho_c$). We have calculated the mass and thickness of the total pasta layers in the inner crust using all the models considered in this work. The parameter sets with larger/smaller symmetry energy and slope parameter estimate thinner/thicker crust and thickness of the pasta structures. Alternatively, a larger negative/positive $K_{\rm sym}$ value corresponds to the thicker/thinner crust and pasta mass and thickness. The pasta mass and thickness are also in agreement with various theoretical constraints. Additionally, rod and slab configurations occupy the largest mass and thickness in the inner crust. The E-RMF models that predict the existence of only two nonspherical pasta phases before transiting into the homogeneous core have the mass and thickness of the slab phase lesser than the rod phase.
   
Quasiperiodic oscillations in soft gamma-ray repeaters are one of the observational means to constrain the inner crust structure and the amount of pasta structures in it. In view of this, we calculate the shear modulus and shear speed in the inner crust of a neutron star by using different methods for the spherical and pasta layers. These quantities are also model-dependent, and considerable uncertainty exists between them. We then consider the pasta layers to have zero shear modulus and neglect the superfluid and entrainment effects to approximate the frequency of fundamental torsional oscillation mode in the crust for the maximum mass. The pasta structure significantly impacts the fundamental frequency mode. Out of 13 EoSs, only two parameter sets, FSUGarnet and IUFSU, agree with the 18Hz observational frequency. Finally, we calculate various neutron star properties for the constructed unified equation of states. The pasta phases do not impact the star's moment of inertia significantly. The fractional crustal moment of inertia ($I_{\rm crust}/I$) for maximum mass estimated from these EoSs are consistent with the pulsar glitch catalog. 

In conclusion, we provide a comprehensive treatment of nuclear pasta properties using the simplistic treatment, and their implication on neutron star properties is analyzed. The unified EoSs constructed in this work stands various theoretical and observational tests and are found suitable for the in-depth investigation of different crust mechanisms. 
\bibliography{pasta}
\end{document}